\documentclass{article}
\usepackage{graphicx}
\usepackage{amssymb}

\begin{document}

\title {Effect of soft and hard x-rays on shock propagation, preheating and ablation characteristics in pure and doped Be ablators}

\author{Karabi Ghosh\footnote{Corresponding author(email: karabi@barc.gov.in)}, Gaurav Mishra\\ \\ Theoretical Physics Section,\\
 Bhabha Atomic Research Centre, Mumbai 400085, India}

\maketitle

\begin{abstract}
In this paper, we analyze the performance of pure and doped Be ablators used for Inertial Confinement Fusion (ICF) pellets in terms of shock velocity, shock breakout temperature, preheat temperature and mass ablation rate through radiation hydrodynamic (RHD) simulations. For this study, we apply a constant radiation profile (drive temperatures varying from 120 - 200 eV) consisting of a low frequency Planck spectrum (soft x-rays) and a high frequency Gaussian spectrum (hard x-rays, commonly termed as "M-band") on a planar foil of the ablator. The fraction of energy density in the hard x-ray spectrum ($\alpha$) has been varied from 0 to 0.25. In pure Be, at lower drive temperatures, the shock velocities are found to rise slowly with $\alpha$. Beyond 170 eV, the shock velocities do not change with $\alpha$. The predominant effect of hard x-rays is to preheat the ablator ahead of the shock front. Steady rise in preheat temperature and shock breakout temperature is observed on increasing the fraction of hard x-rays. Also, mass ablation rates in Be are found to be unaffected by hard x-rays. Preheating can be mitigated by doping Be with a mid-Z element Cu as its opacity is much higher in the high frequency region. On doping Be with 1\% Cu, the shock velocities decrease slightly compared to pure Be. However, higher shock velocities are observed on increasing the fraction of M-band. Similar to pure Be, preheat and shock breakout temperatures have a strong dependence on $\alpha$. We observe significant decrease in shock breakout and maximum preheat temperature in doped Be foil. Steady rise in these temperatures is observed on increasing $\alpha$. Unlike pure Be, the mass ablation rate is found to increase slowly with M-band fraction in Cu doped Be. We have proposed new scaling relations for shock velocity, shock breakout temperature, maximum preheat temperature and mass ablation rate with the radiation temperature and the fraction of M-band energy density in both pure and doped Be ablators. In terms of ablator performance, Cu doped Be ablator is found to be superior to pure Be. Though doping significantly reduces preheating, the mass ablation rates are nearly unaffected.

\end{abstract}


\section{Introduction}

In indirect drive Inertial Confinement Fusion (ICF) \cite{Atzeni,Lindl2004}, laser beams are focussed on the inner walls of a hohlraum through laser entrance holes. Due to multiple absorptions and reemissions, nearly isotropic distribution of x-rays is established within the hohlraum. These x-rays ablate the outer layers of the fusion capsule kept at the centre, thereby causing compression, ignition and burn. Generally, the radiation field inside the hohlraum is assumed to be a blackbody spectrum (Planckian distribution). However, experiments and radiation hydrodynamic (RHD) simulations suggest that the blackbody approximation of the hohlraum radiation field provides an incorrect calculation of the shock temperature and breakout time at the rear surface of a Cu doped Be foil. RHD simulations involving proper accounting of hard x-rays in the time dependent radiation spectrum of hohlraum leads to experimentally agreeable results \cite{olson2004}. In reality, ICF experiments employing  Au hohlraums always consist of radiation spectrum that includes photon energies in the range of 2-5 keV (Au M-band) along with low energy x-rays. The M-band originates in and near the hot ($\sim$ 1-5 keV), low density ($n_e< 10^{22}cm^{-3}$) coronal plasma in which the laser light is absorbed when the laser strikes the high-Z hohlraum wall \cite{olson2004}. The correct modelling of M-band spectrum in RHD simulations requires its location and magnitude in the total x-ray spectrum corresponding to various hohlraum temperatures. The detailed spectrometer analysis of M-band spectrum suggests various  broad peaks at around 2.1, 2.5, 2.8 and 3.3 keV \cite{Robey}. M-band fraction can be upto 15\% of the total hohlraum X-ray flux emitted \cite{Troussel}. Using the SGIII-prototype-laser facility, the hohlraum peak radiation temperature $T_r$ was $\sim$ 195 eV and the M-band fraction $f_m$ approached 10\% \cite{LiHEDP}. M-band fraction has been predicted to be 9\% in a hohlraum driven by a 1 ns flattop pulse of 6 kJ laser energy using the shock wave technique \cite{POP21}. It has also been predicted based on experimental results that at 1.3 MJ in an ignition hohlraum, a peak x-ray drive of 14.3 TW/sr is obtained at 305 eV with a M-band fraction of 20\% \cite{POP18}. The radiant power and spectral distribution of the radiation of the Au plasma has been measured using a broadband X-ray spectrometer called DMX on the Omega laser facility at the Laboratory for Laser Energetics (LLE) \cite{Troussel}. In the SG-III prototype laser facility, a flat-response X-ray detector (F-XRD) and M-band X-ray detector (M-XRD) were used to measure the X-ray fluxes in the range of 0.1 - 4 keV and 1.6 - 4.4 keV respectively.  $T_r$ and $f_m$ have also been measured simultaneously using the shock wave technique. Both Al and Ti have been used as the witness plates mounted on the hohlraum wall and the shock velocities measured. For a pure Planckian radiation source, the peak radiation temperature $T_r$ can be determined via a scaling relation of $T_r$ with the shock velocity $U_s$. In a hot Au hohlraum, the M-band fraction cannot be ignored, and $U_s$ varies with both $T_r$ and $f_m$. As $U_s$ decreases with $f_m$ for Al but increases in Ti, both $T_r$ and $f_m$ are obtained simultaneously from the crossing of the contour lines. The values of $f_m$ obtained from shock wave technique are the maximums, while those obtained from SXS are their temporal behaviour. Thus, the results of the two techniques are complementary \cite{POP18AlTi}. By dividing the incident radiation source into a low frequency Planckian and a high frequency Gaussian distribution, the shock velocities can be obtained through radiation hydrodynamic simulations and the M-band fraction can be obtained along with the radiation temperature. Analyzing the motion of a sample due to preheat has also been used to provide information on the fraction of soft and hard x-rays in the source \cite{POP20Al}. To determine the rear surface spectral temperature, the spectral intensity averaged over a 200 ps window at the time of shock breakout was compared to a Planck form. The unloading material was assumed to radiate locally as a blackbody \cite{JAP58}. The temperature attained by a planar target due to shock heating is determined by the rear surface luminiscence at the time the shock emerges. These measurements underestimate the true shock heated material temperature because of the rarefaction of the material due to shock breakout and the finite measurement time required in the experiment \cite{OC53}.

 The high energy x-rays pass through the ablator ahead of the shock front because of their longer mean free paths. This leads to preheat of the fusion fuel and degrades the implosion characteristics \cite{preheat-Olson-2003}. This would decrease the capsule compressibility and result in lower areal density. Preheating ahead of the shock front also causes significant shock propagation variation in the capsule thereby reducing the final yield \cite{POP26-2019}. As the rear surface of the foil expands due to preheating, large uncertainties in EOS diagnostics creep in \cite{POP20Al}. Preheating also makes it difficult to maintain a low adiabat in ICF capsules thereby requiring much higher implosion kinetic energy \cite{POP23-Cheng}. It may be noted that the presence of x-rays in the M-band  also affects the shock velocities in mid-Z elements like Al and Ti. Moreover, a scaling relation between drive temperature and shock velocity is also proposed for the above mentioned elements \cite{POP18AlTi}. Though the impact of M-band on degradation of implosion characteristics, instability, yield, etc., have been studied well, its effect on preheat temperature, shock temperature, mass ablation rate, etc. in low-Z ablators have not been analyzed in detail. The effect of hard X-rays on shock velocity has been quantitatively studied for mid-Z elements \cite{POP18AlTi}. No such quantitative study has yet been performed for low-Z ablators. Also, the variation in shock temperature due to the presence of hard x-rays has not been investigated earlier. Preheating of material ahead of the shock front because of hard x-rays have been observed for low-Z ablators. A relationship between the absorbed energy and preheat temperature has been obtained in a planar plastic (CH) foil \cite{POP22thin}. Nevertheless, the dependence of preheat temperature on drive temperature and $\alpha$ has never been established. The mass ablation rate, which decides the ablator performance, has been scaled with the radiation temperature \cite{POP18Olson}. However, its variation with the fraction of M-band energy density is missing. 

  The most common ablators used for fusion ignition experiments on the National Ignition Facility (NIF) are beryllium, plastic and high-density carbon or synthetic diamond \cite{NF-2004}. The opacities of the ablators being low for high frequency photons, the hard x-rays preheat the DT fuel to an unacceptable degree \cite{POP-2010-Clark}. In order to enhance the capsule performance, doped ablators like beryllium doped with copper, plastic (CH) doped with germanium or silicon, nano-crystalline synthetic diamond, also known as high density carbon, doped with tantalum are preferred \cite{POP-1996,POP-1998_1953,POP-1998_3708}. As Be has the lowest opacity, highest density, lowest specific heat, highest tensile strength, and highest thermal conductivity, it is more stable hydrodynamically than CH and hence is the most preferred ablator material \cite{olson2004},\cite{POP-2010-Clark}. For our RHD simulations, we have used pure Be and Cu doped Be foils. The concentration of the dopant (Cu) is optimized to 1\% as too little dopant cannot prevent preheating, whereas too much may reduce the implosion velocity \cite{olson2004}.
  
The main aim of this work is to perform quantitative analysis of the effect of soft and hard x-rays on shock velocity, shock breakout temperature, maximum preheat temperature and mass ablation rate in Be and Be +1\%Cu ablator foils. We have proposed new scaling laws for these quantities. In addition to providing information about drive temperature and M-band fraction of an unknown radiation source, these scaling relations can be used to validate new radiation hydrodynamic codes. Our results also demonstrate the advantages of doping Be with a small amount of mid-Z element Cu.  

This paper is organized as follows: In Section \ref{SimuModel}, the RHD simulation model has been described in brief. The programs used for generating EOS and opacity of pure and doped Be ablators have also been introduced. We have described the incorporation of high frequency Gaussian source in the Planck spectrum to analyze the effect of M-band in the simulation. The simulation results are presented in Section \ref{res}. Based on the physics of propagation of strong shocks, scaling relations for shock velocity, shock breakout temperature, preheat temperature and mass ablation rate have been obtained. Finally, important conclusions have been presented in Section \ref{conc}.   

\section{Simulation model}\label{SimuModel}
\begin{figure}    
   \begin{center}
       \includegraphics[width=0.8\linewidth]{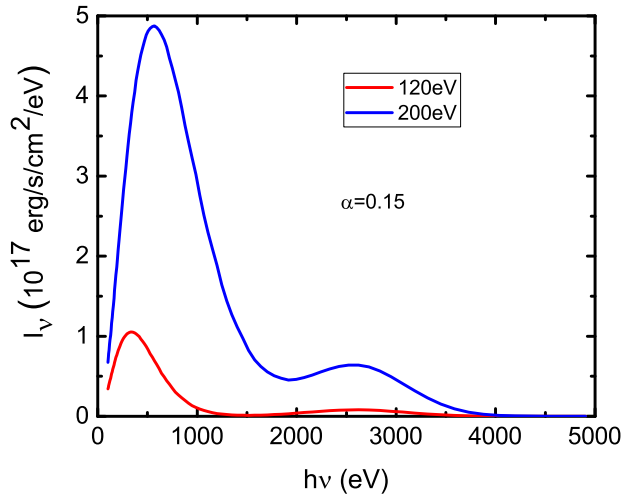}
      \caption{Spectrum of the radiation drive for two radiation temperatures, viz., 120 eV and 200 eV for $\alpha$ = 0.15.}
      \label{source}
  \end{center}
\end{figure}

The one-dimensional multigroup RHD code MULTI \cite{ramis1988multi} has been used for simulation. Opacity of the ablators and doped ablators have been generated using AALS \cite{HEDP2019} which uses the screened hydrogenic model of More with l-splitting. In this simulation, the X-ray photons with energies ranging from 100 eV to 5keV have been divided into 100 groups \cite{HEDP2019}. Equation of state (EOS) has been obtained from QEOS based code TFEOS \cite{HEDP2019,more1988new,MishraHEDP2018}. A constant radiation drive is applied on one side of a planar foil. The strength of the temperature drive is varied from 120 eV to 200 eV. The X-ray source consists of an equilibrium Planckian distribution along with a high frequency Gaussian distribution centred at 2.6 keV and FWHM of 500 eV \cite{Li_POP-2015,preheatbarc}. $\alpha$ is the fraction of the total energy density due to the Gaussian distribution and is defined as 
\begin{equation}
\alpha=\frac{U_m}{U_p}
\end{equation}
with the total energy density in the Gaussian distribution given by $U_m=\int_{0}^{\infty}U_{\nu m}dT_\nu$. The Gaussian spectral energy density is given by
\begin{equation}
U_{\nu m}=\frac{\alpha U_p}{\sqrt{2\pi \sigma}}exp\frac{-(T_\nu-T_m)^2}{2\sigma^2}
\end{equation}
Total energy density of Planckian radiation field ($U_p$) is
\begin{equation}
U_p=\frac{8\pi^5 (k_B T_R)^4}{15c^3h^3}
\end{equation} 
Also, $f_m$ represents the fraction of X-rays in the M-band i.e., 2 to 5 keV. The M-band fraction ($f_m$) consists of the Gaussian and the high energy part of the Planckian distribution and is defined as 
\begin{equation}
f_m=\frac{\int_{2keV}^{5keV}I_\nu dT_\nu}{\int_{0}^{\infty}I_\nu dT_\nu}
\end{equation}
$f_m$ and $\alpha$ can be related as 
\begin{equation}
f_m=\frac{\int_{2keV}^{5keV}(1-\alpha)I_{\nu p} dT_\nu}{\int_{0}^{\infty}I_\nu dT_\nu}+\frac{\int_{2keV}^{5keV}I_{\nu m} dT_\nu}{\int_{0}^{\infty}I_\nu dT_\nu}
\end{equation}
We have written a supportive program for calculating $f_m$ for a given value of $\alpha$ and a constant drive temperature ($T_r$) \cite{noneqbarc}. A fitting function of the form $f_m=p+q \alpha +r T_r$ has been obtained . The values of the fitting coefficients p, q and r are found to be -0.01269, 0.88206 and 9.6981 $\times 10^{-5}$ for $T_r$ varying from 120 to 200eV and $\alpha$ ranging from 0 to 0.25. We have restricted the value of $\alpha$ to 0.25 as higher contribution of M-band has not been observed experimentally \cite{POP18}. We have modified the RHD code MULTI to incorporate the non-equilibrium Gaussian source in addition to the low-frequency Planckian spectrum \cite{noneqbarc}. The fraction of X-ray energy density in the M-band ($\alpha$) is fed as input to the code. Hence, we have determined the variation in thermodynamic parameters like shock velocity, temperature, mass ablation rate, etc. in terms of $\alpha$ and not $f_m$. However, $f_m$ is the experimentally measurable quantity. For any value of $\alpha$, the value of $f_m$ can be obtained for a constant radiation temperature using the above fitted formula relating $f_m$ and $\alpha$. A similar formula relating $f_m$ and $\alpha$ for a radiation temperature of 300 eV has been obtained in ref. \cite{Jiang}. A representative plot of the spectral intensity ($I_\nu$) is shown in Fig. \ref{source} for $\alpha$ = 0.15 at drive temperatures of 120 eV and 200 eV, respectively. For a radiation drive of 120 eV, separate Planck and Gaussian spectrums are observed below and above the photon energy of 1500 eV respectively. However, at a higher drive temperature of 200 eV, as the peak of the Planckian spectrum shifts to higher frequencies, overlap between the Gaussian and Planckian spectrum is observed. 

\section{Results}\label{res}
\begin{figure}    
   \begin{center}
       \includegraphics[width=0.8\linewidth]{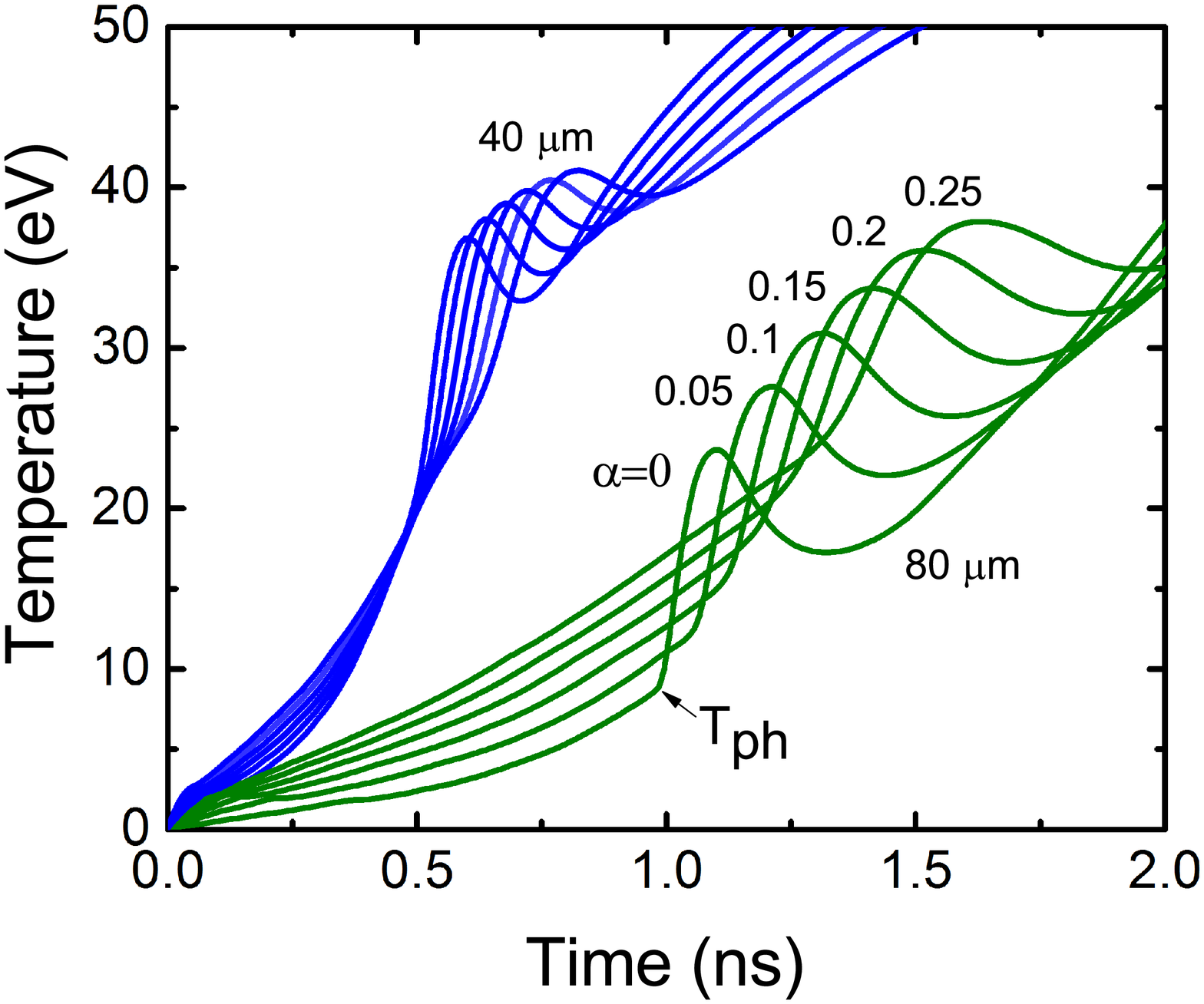}
      \caption{Variation of temperature with time at the rear surface of a Be foil of thickness 40 $\mu m$ and 80 $\mu m$ for $\alpha$ varying from 0 to 0.25. The incident radiation temperature is 200 eV.}
      \label{40-80}
  \end{center}
\end{figure}

\begin{figure}    
   \begin{center}
       \includegraphics[width=0.8\linewidth]{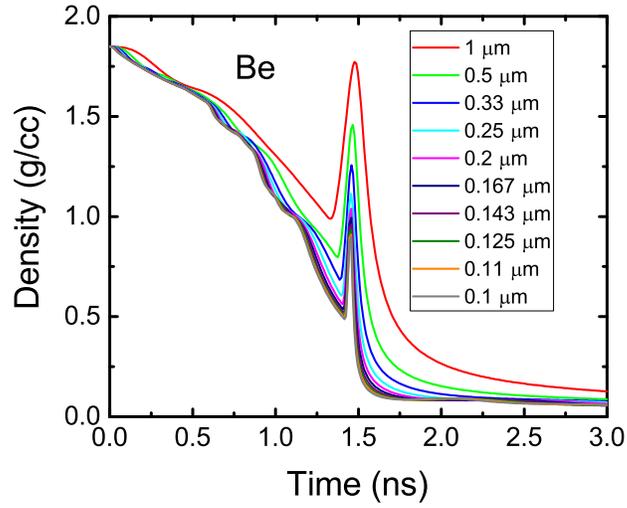}
      \caption{Variation of density ($\rho$) with time for $T_r$=160 eV and $\alpha$=0.}
      \label{dens-conv}
  \end{center}
\end{figure}

\begin{figure}    
   \begin{center}
       \includegraphics[width=0.8\linewidth]{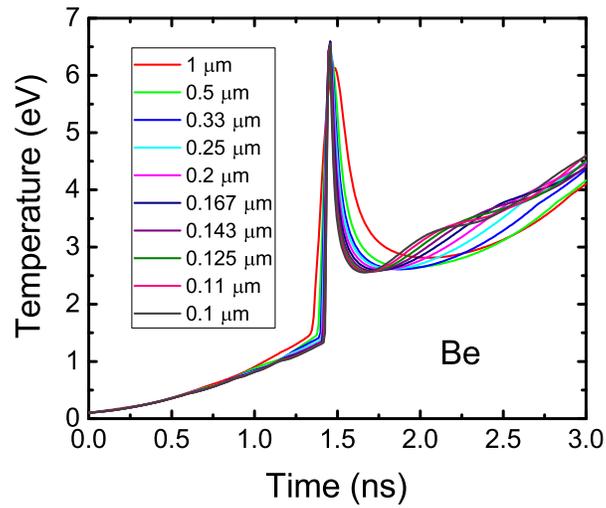}
      \caption{Variation of temperature ($T$) with time for $T_r$=160 eV and $\alpha$=0.}
      \label{temp-conv}
  \end{center}
\end{figure}

For the purpose of simulation, we choose a Be foil of thickness 80 $\mu m$ \cite{olson2004}. If a thinner foil of 40 $\mu m$ is chosen, the foil ablates before a stable shock is formed. In Fig. \ref{40-80}, we plot the electron temperature at the rear surface as a function of time for 40 $\mu m$ and 80 $\mu m$ thick foils at 200 eV. For the 40 $\mu m$ thick foil, the electron temperature increases rapidly due to preheating. Hence, we do not observe the arrival of a steep shock. As a result, the determination of a maximum preheat temperature just before the arrival of the shock is not posssible. However, for a 80 $\mu m$ foil, preheating at the rear surface is slow. A steep shock front is formed for $\alpha$ varying from 0 to 0.25. In this case an accurate determination of the maximum preheat temperature is easy. We need not choose a thicker foil as larger simulation time is necessary. For optimized thickness of 80 $\mu m$, our RHD simulations give maximum preheat temperature of about 2 eV for 120 eV drive temperature with $\alpha$ equal to 0.05 and 25 eV for 200 eV drive temperature with $\alpha$ equal to 0.25. Further, the choice of mesh width for a given thickness needs special attention so as to clearly resolve  the various parameters like density and temperature at rear surface of the foil \cite{Bradley}. For this, we have carried out a convergence study on temporal variation of these parameters at rear surface by varying the mesh widths in the range of 1 $\mu m$ to 0.1 $\mu m$. The results are shown in Figs. \ref{dens-conv} and \ref{temp-conv} for Be foil illuminated by Planckian radiation drive of temperature 160 eV. We note from these plots that the results converge for a mesh width of 0.2 $\mu m$. So the same mesh width is used for all simulations carried out in the present study.

\subsection{Shock velocity scaling}
The inner walls of ICF hohlraums are coated with high-Z elements which absorb the laser energy and convert them to x-rays via reemission. If the x-rays from the hohlraum are allowed to fall on an ablator foil, the low-Z ablator absorbs the radiation and  a shock wave is generated into the medium. For a Planckian radiation source, the material ahead of the shock front is at rest, whereas, the medium through which shock has moved acquires a particle velocity ($U_p$). The shock velocity ($U_s$) is obtained from a linear fit of shock position versus time in the planar foil \cite{MishraHEDP2018}.  

For a Planckian blackbody spectrum, the radiation temperature can be related to the shock velocity as [\cite{Atzeni},\cite{MishraHEDP2018},\cite{Hatchett}]
\begin{equation}\label{TrUs}
T_r=\eta U_s^\phi.
\end{equation}
where the coefficients $\eta$ and $\phi$ vary from one material to the other. We arrive at this relation by treating x-ray driven ablation as a stationary process. Ablation pressure $P \sim (1-\alpha_d)T_r^4/U_p \sim (3/4)\rho U_s^2$. Here, $\alpha_d$ is the albedo and $\rho$ is the density of the ablator. The ablated particle speed is given by $U_p \sim \sqrt{Z_0T_r/A}$. Here, $Z_0$ is the charge state and A is the mass number of the ablator. The shock velocity can be experimentally measured using a wedge shaped foil. The shock breaks out at different times along the wedge and emits radiation because of shock heating.

However, when the incident radiation spectrum is not a Planckian, but consists of a high frequency Gaussian spectrum as well, the coefficients $\eta$ and $\phi$ are no more constant but depend upon $f_m$ or $\alpha$ for a given material \cite{POP18AlTi}. This happens because in a non-Planckian spectrum, the high energy photons propagate ahead of the shock front and cause preheating of the material. Therefore, the shock moves into a hot and expanded medium. Thus the shock speed is found to vary not only with the radiation temperature, but also with the fraction of radiation energy densities at higher frequencies ($\alpha$). 
In that case, Eq. \ref{TrUs} modifies to following relation,
\begin{equation}\label{TrUsa}
T_r=\eta(\alpha)U_s^{\phi(\alpha)}.
\end{equation}

\begin{figure}    
   \begin{center}
       \includegraphics[width=0.8\linewidth]{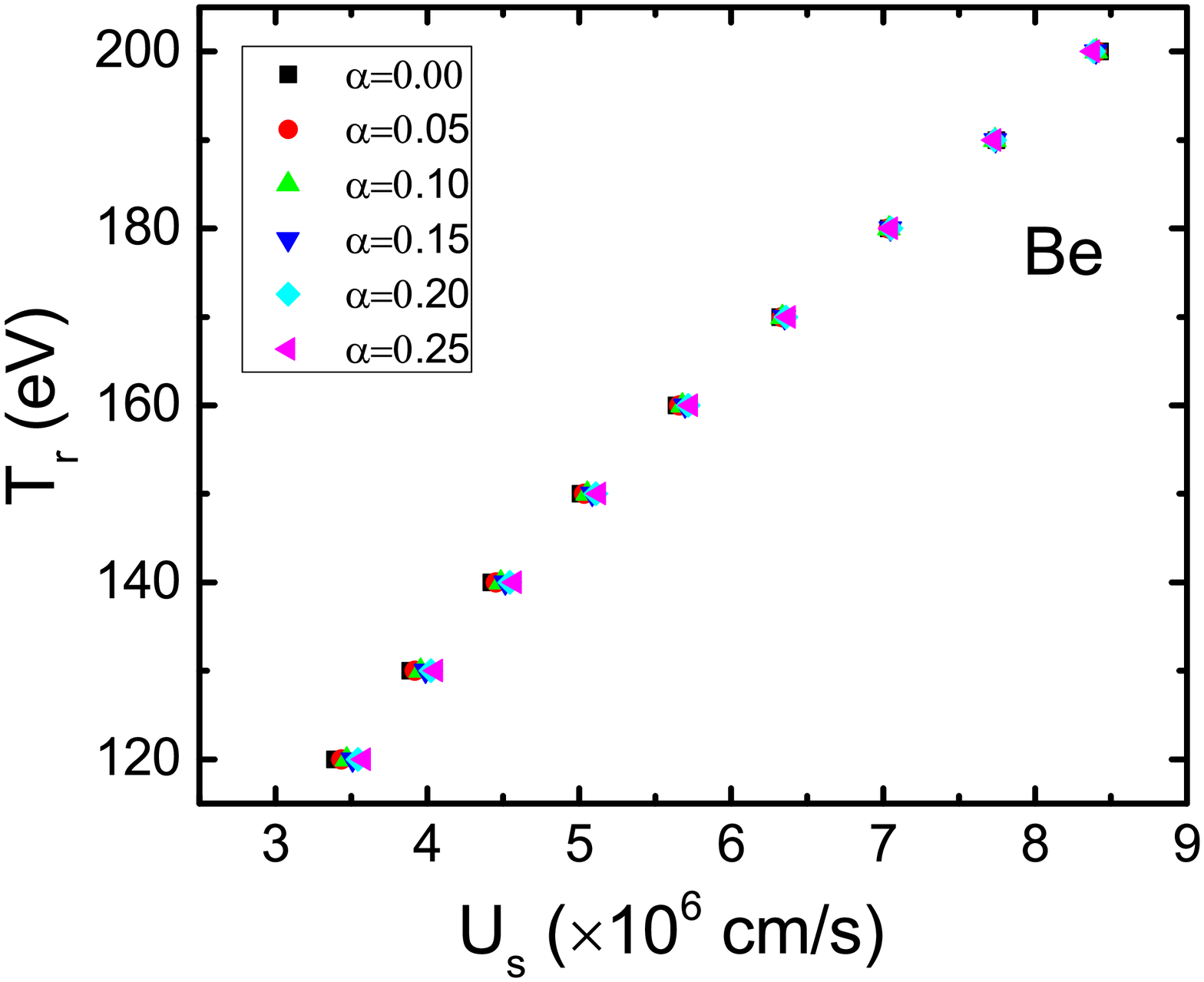}
      \caption{Variation of $T_r$ with $U_s$ in a Be foil for values of $\alpha$ varying from 0 to 0.25.\label{TRvsUs_Be}}
  \end{center}
\end{figure}

\begin{figure}    
   \begin{center}
       \includegraphics[width=0.8\linewidth]{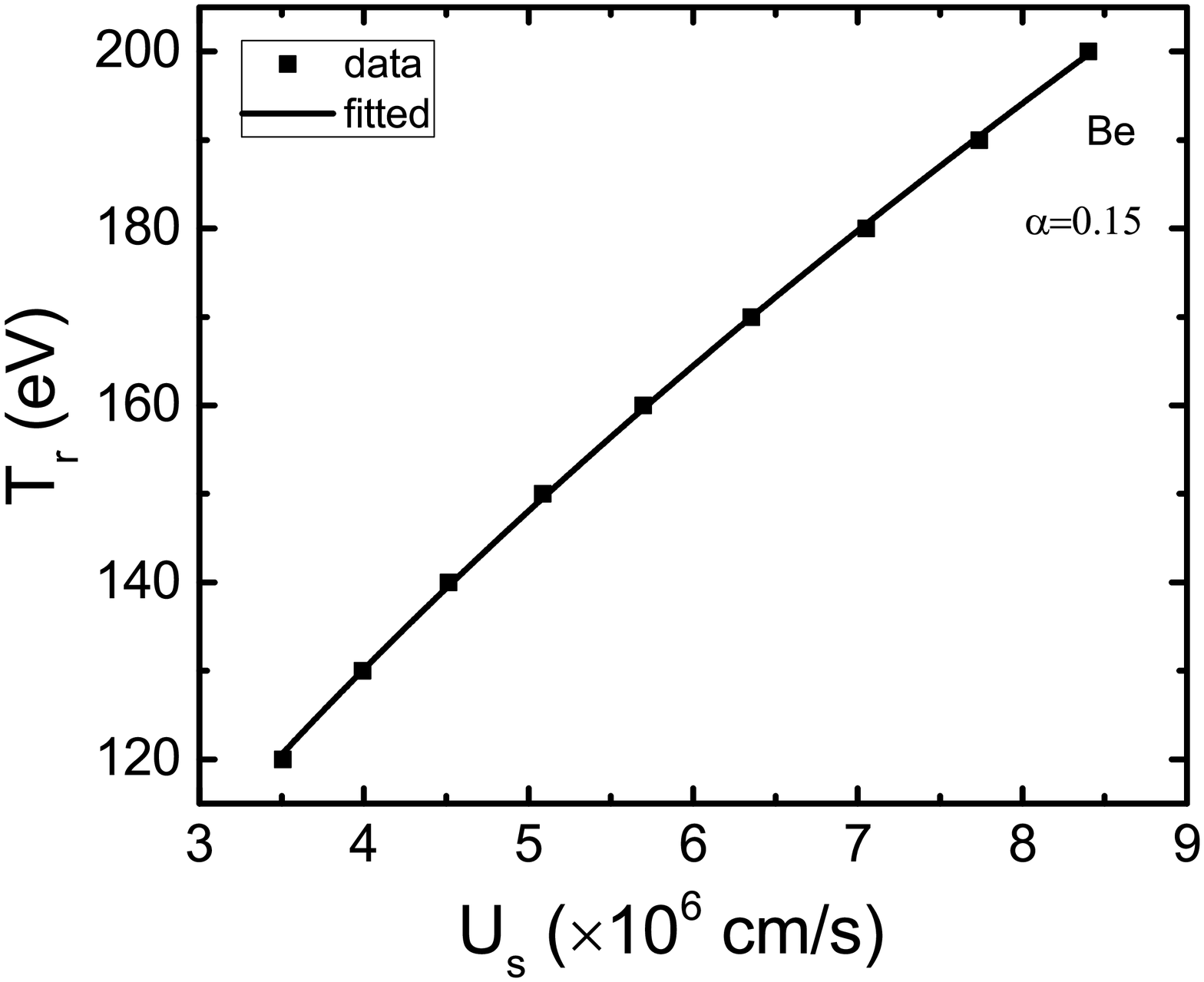}
      \caption{Fitting of $T_r$ with $U_s$ in a Be foil for $\alpha$ = 0.15.\label{TRvsUs_Be_fit}}
  \end{center}
\end{figure}

\begin{figure}    
   \begin{center}
       \includegraphics[width=0.8\linewidth]{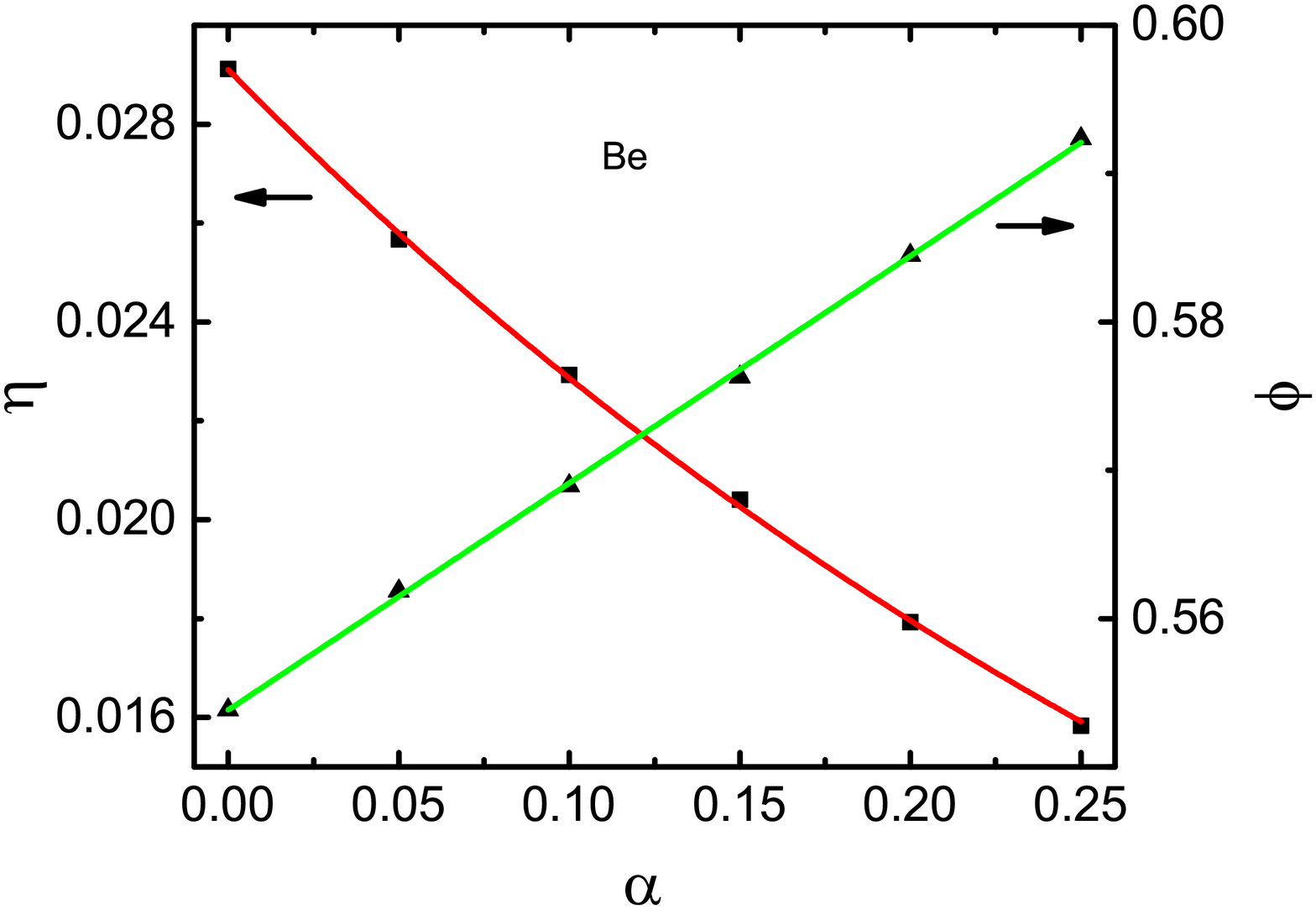}
      \caption{Variation of $\eta$ and $\phi$ with $\alpha$ in a Be foil.\label{eta_phi_alpha_Be}}
  \end{center}
\end{figure}

\begin{figure}    
   \begin{center}
       \includegraphics[width=0.8\linewidth]{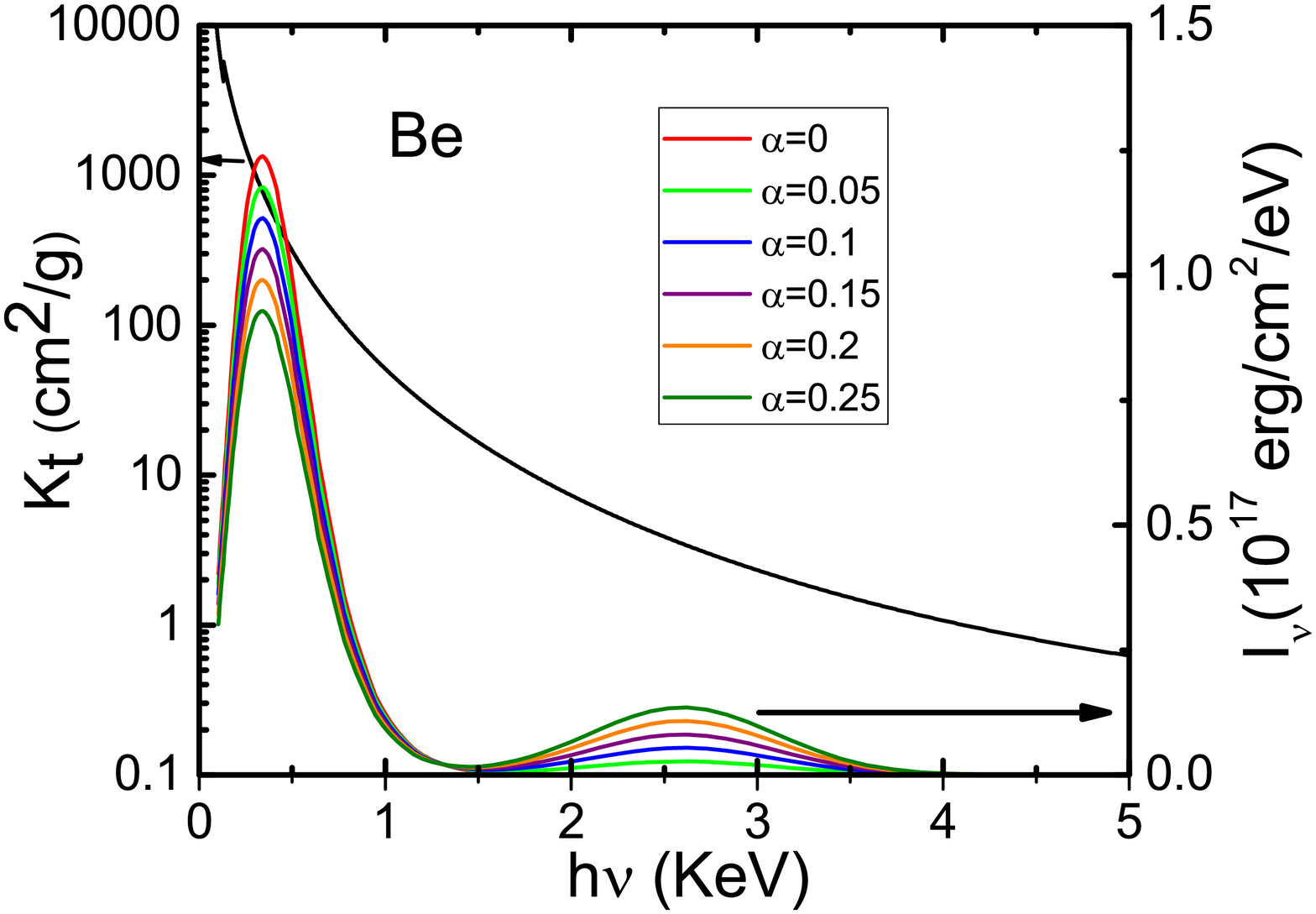}
      \caption{Variation of total extinction coefficient of Be with photon energy at a temperature of 120eV. The incident intensities are also plotted in the same graph for $\alpha$ varying from 0 to 0.25.\label{KT-I_Be-120}}
  \end{center}
\end{figure}

\begin{figure}    
   \begin{center}
       \includegraphics[width=0.8\linewidth]{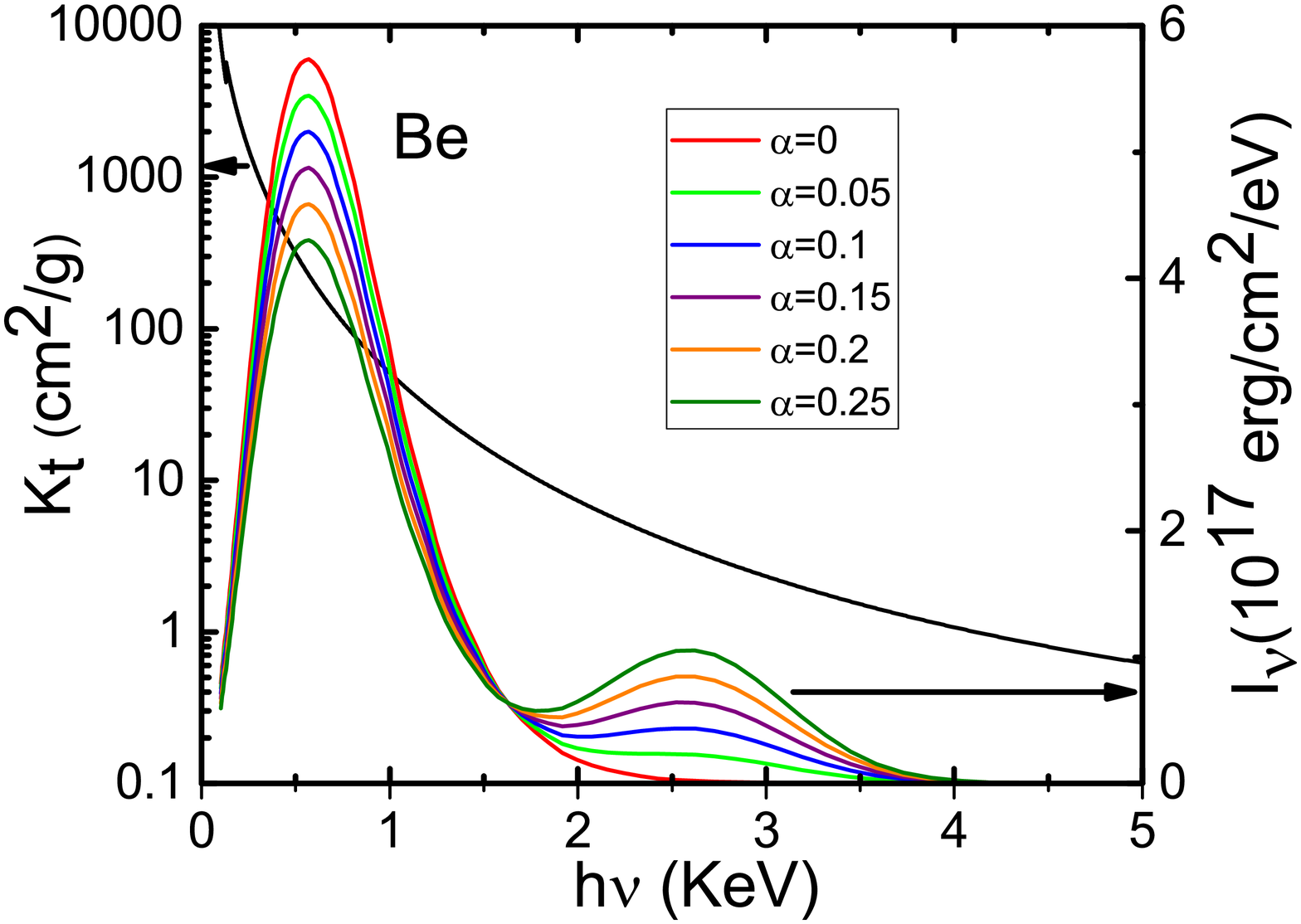}
      \caption{Variation of total extinction coefficient of Be with photon energy at a temperature of 200eV. The incident intensities are also plotted in the same graph for $\alpha$ varying from 0 to 0.25.\label{KT-I_Be-200}}
  \end{center}
\end{figure}

In Fig. \ref{TRvsUs_Be}, we plot the data for radiation drive temperature $T_r$ as a function of shock velocity $U_s$ in pure Be foil for different values of $\alpha$. We have also fitted the data for $T_r$ Vs $U_s$ using Eq. \ref{TrUsa} for all $\alpha$ values.  A representative fit is shown in Fig. \ref{TRvsUs_Be_fit} for $\alpha$ = 0.15. As depicted by Eq. \ref{TrUsa}, the dependence of $\eta(\alpha)$ and $\phi(\alpha)$ on different values of $\alpha$ is shown in Fig. \ref{eta_phi_alpha_Be}. We find that $\eta$ follows a power law dependence on $\alpha$, viz. $\eta(\alpha)=a.b^\alpha$ whereas $\phi$ is found to vary linearly with $\alpha$ as $\phi(\alpha)=c+d.\alpha$. This form of shock velocity scaling has been proposed by Li et. al. and successfully applied to Al and Ti foils \cite{POP18AlTi}. The values of a,b,c and d are 0.0291, 0.08951, 0.5538 and 0.1531 respectively for Be. It may be noted from Fig. \ref{TRvsUs_Be} and Eq. \ref{TrUsa} that shock velocity increases with temperature for a given value of $\alpha$. We further observe that shock velocity increases with $\alpha$ for low drive temperatures whereas it becomes nearly independent of $\alpha$ at higher drive temperatures. To explain this, we have plotted the total extinction coefficient of pure Be as a function of frequency in Fig. \ref{KT-I_Be-120} at a temperature of 120 eV and density of 2 g/cc. In the same graph, we have also shown the incident intensities for various $\alpha$ values at 120 eV. The total extinction coefficient of Be is found to fall very sharply with frequency. The frequency at which the Planckian spectrum peaks, the total extinction coefficient is as high as 760 $cm^2/g$, but falls to 3.4 $cm^2/g$ at the peak of the Gaussian spectrum.  As $\alpha$ increases, the intensity at the peak of the Planckian reduces and that of the Gaussian rises. We observe from Fig. \ref{KT-I_Be-120} that the total extinction coefficient of the medium is higher for low energy photons compared to those having higher energies. So, we can define the effective opacity as the opacity encountered by all energy photons. For higher $\alpha$, i.e., more high energy photons, the effective opacity reduces as the total extinction coefficient is low at higher energies. Thus, with increase in $\alpha$ at 120 eV, the effective opacity reduces significantly. At a fixed temperature, the shock velocity $U_s\propto (1-\alpha_d)$, where $\alpha_d$ is the albedo of the foil. The albedo is defined as reemitted flux to the incident flux. At a fixed temperature, a material having lower opacity reemits less and hence has a lower value of albedo. The shock velocity being proportional to $(1-\alpha_d)$, increases as a result as shown in Fig. \ref{TRvsUs_Be}. However, we do not observe any significant increase in $U_s$ with $\alpha$ beyond 170 eV. This can be explained as follows: In Fig. \ref{KT-I_Be-200}, we have plotted the total extinction coefficient of pure Be as a function of frequency at a temperature of 200eV and a density of 2 g/cc. In the same graph, we have also shown the incident intensities for various $\alpha$ values at 200 eV. Here, we observe that the Planckian has shifted from 0.34 KeV (at 120eV) to 0.554 KeV. At this frequency the total extinction coefficient has reduced to a low value of 227 $cm^2/g$. Hence, the variation in $\alpha$ makes a negligible change in albedos, and hence no change in shock velocity could be determined \cite{MishraHEDP2018}.

\begin{figure}    
   \begin{center}
       \includegraphics[width=0.8\linewidth]{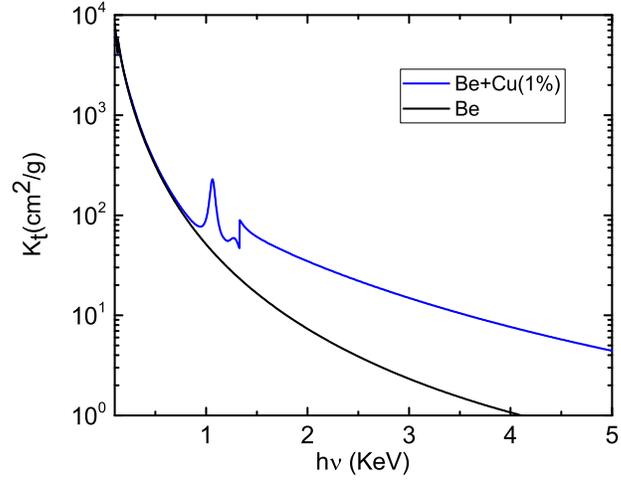}
      \caption{Variation of extinction coefficient of pure Be and Be doped with 1\% Cu as a function of photon energies for a temperature of 120 eV at a density of 2g/cc.\label{KT-BeCu}}
  \end{center}
\end{figure}

\begin{figure}   
\begin{center}
       \includegraphics[width=0.8\linewidth]{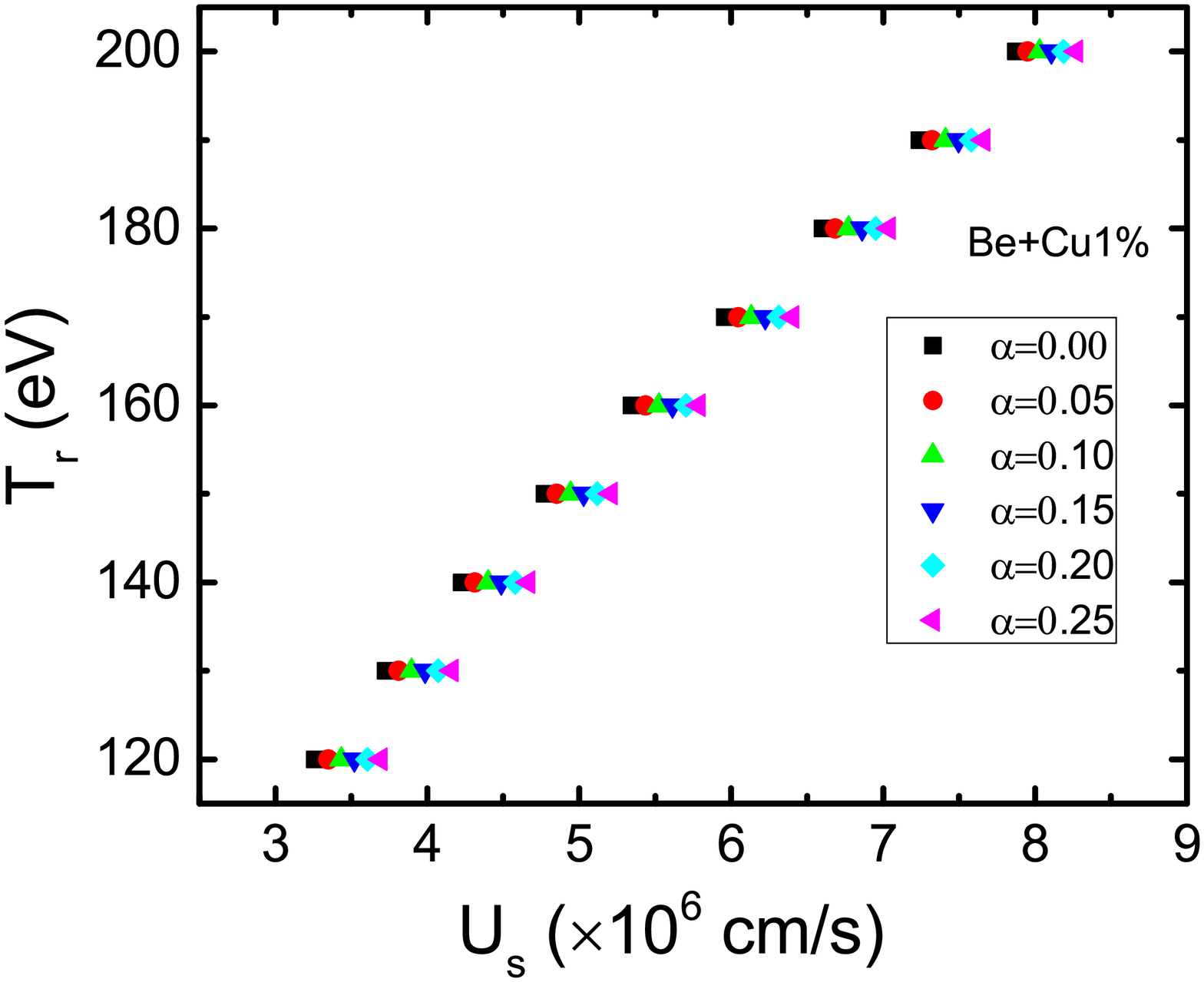}
      \caption{Variation of $T_r$ with $U_s$ in a Be+1\%Cu foil for values of $\alpha$ varying from 0 to 0.25.\label{TRvsUs_BeCu}}
      \end{center} 
\end{figure}

\begin{figure}    
   \begin{center}
       \includegraphics[width=0.8\linewidth]{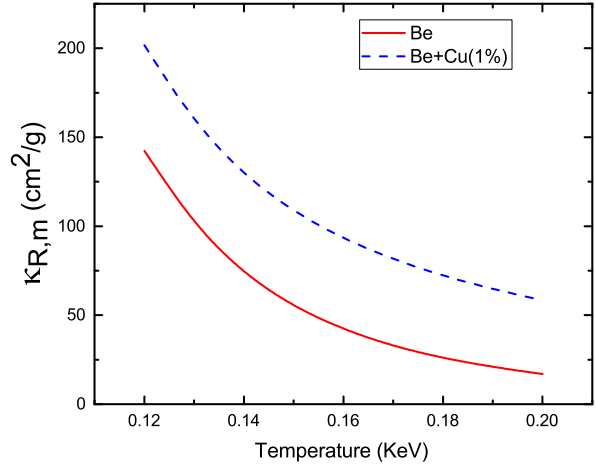}
      \caption{Variation of Rosseland opacity of pure and doped Be ablator with temperature at a density of 2g/cc.\label{op-BeCu}}
  \end{center}
\end{figure}

\begin{figure}    
   \begin{center}
       \includegraphics[width=0.8\linewidth]{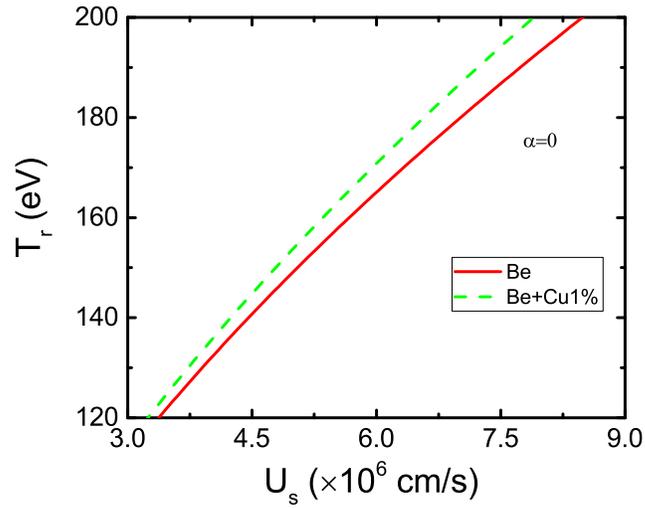}
      \caption{Variation of $T_r$ with $U_s$ for pure Be foil and Be doped with 1\% Cu for $\alpha$ = 0.\label{TRvsUs_a0}}
  \end{center}
\end{figure}

\begin{figure}    
   \begin{center}
       \includegraphics[width=0.8\linewidth]{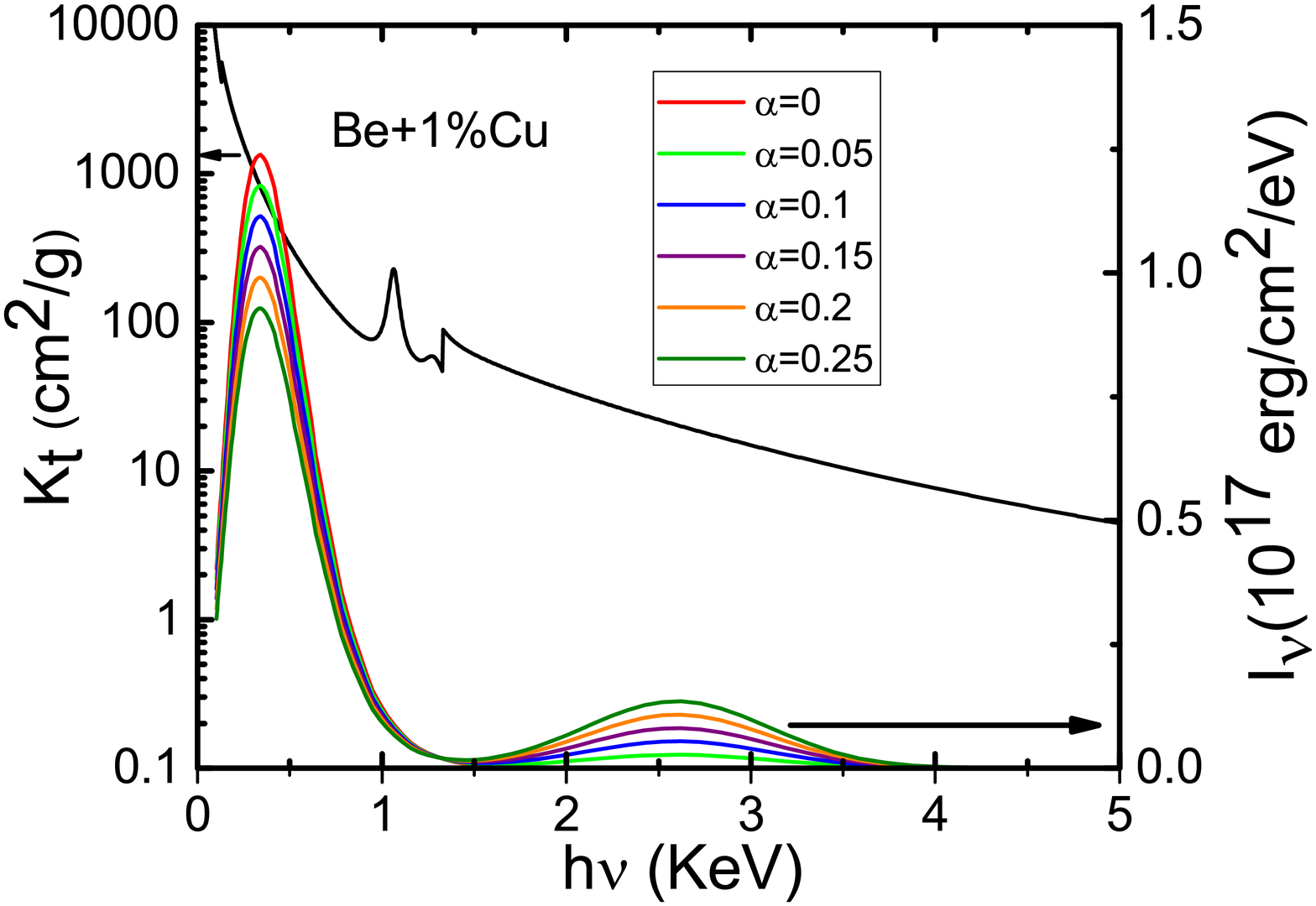}
      \caption{Variation of total extinction coefficient of Be doped with 1\% Cu as a function of photon energy at a temperature of 120eV. The incident intensities are also plotted in the same graph for $\alpha$ varying from 0 to 0.25.\label{KT-I-BeCu1-120eV}}
  \end{center}
\end{figure}

\begin{figure}    
   \begin{center}
       \includegraphics[width=0.8\linewidth]{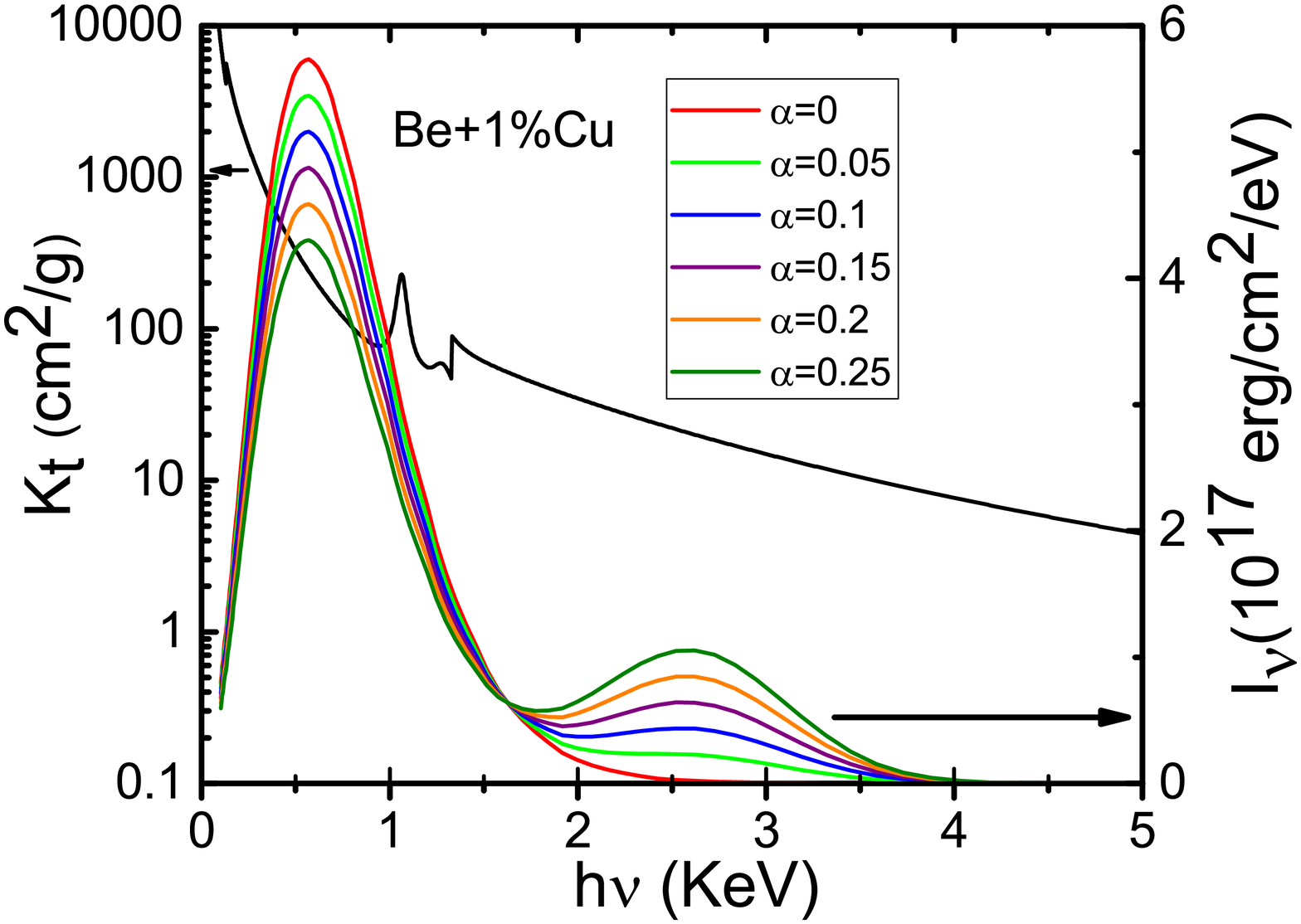}
      \caption{Variation of total extinction coefficient of Cu doped Be with photon energy at a temperature of 200eV. The incident intensities are also plotted in the same graph for $\alpha$ varying from 0 to 0.25.\label{KT-I-BeCu1-200eV}}
  \end{center}
\end{figure}

\begin{figure}    
   \begin{center}
       \includegraphics[width=0.8\linewidth]{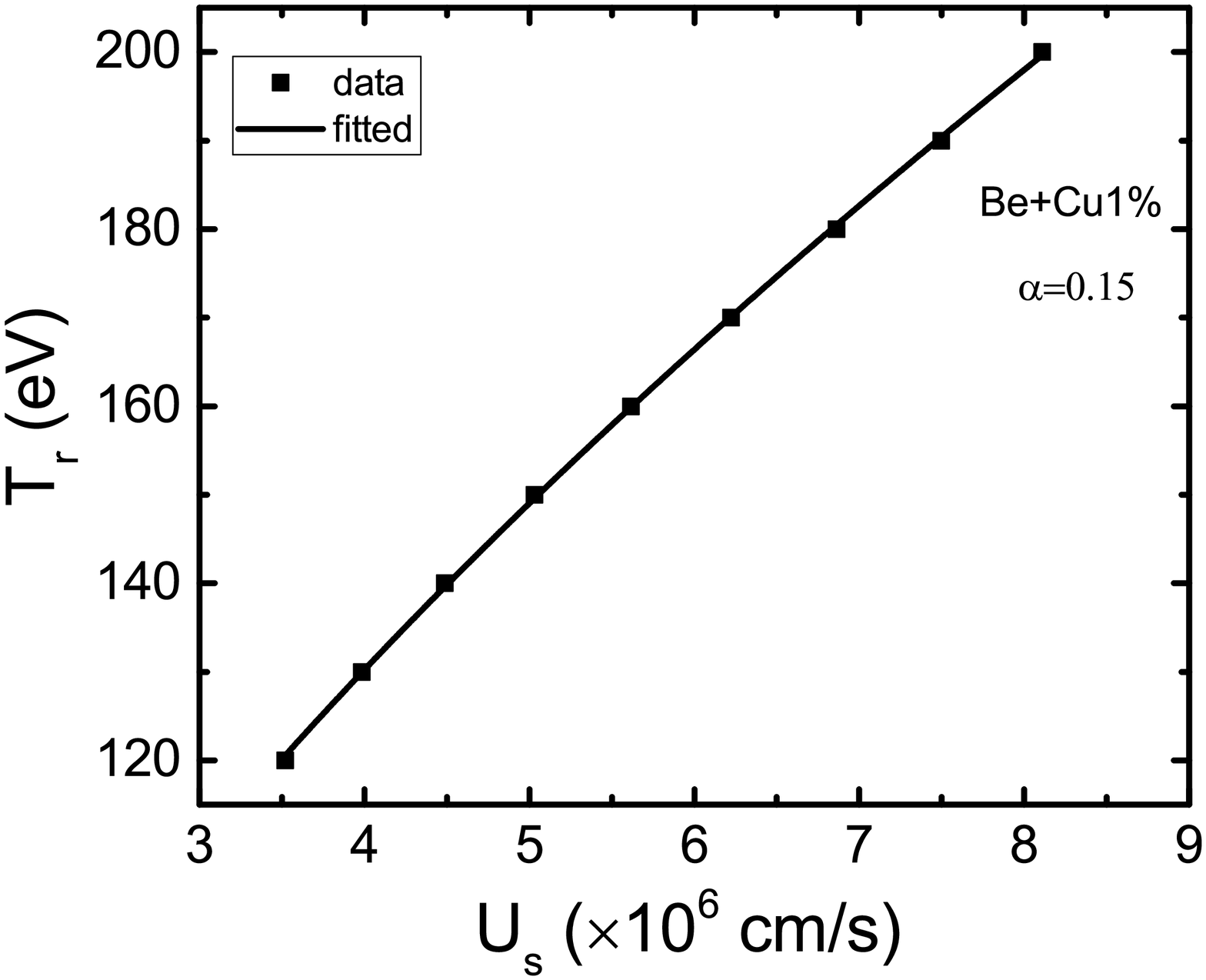}
      \caption{Fitting of $T_r$ with $U_s$ in a Be+1\%Cu foil for $\alpha$ = 0.15.\label{TRvsUs_BeCu1_fit}  } 
  \end{center}
\end{figure}

\begin{figure}    
   \begin{center}
       \includegraphics[width=0.8\linewidth]{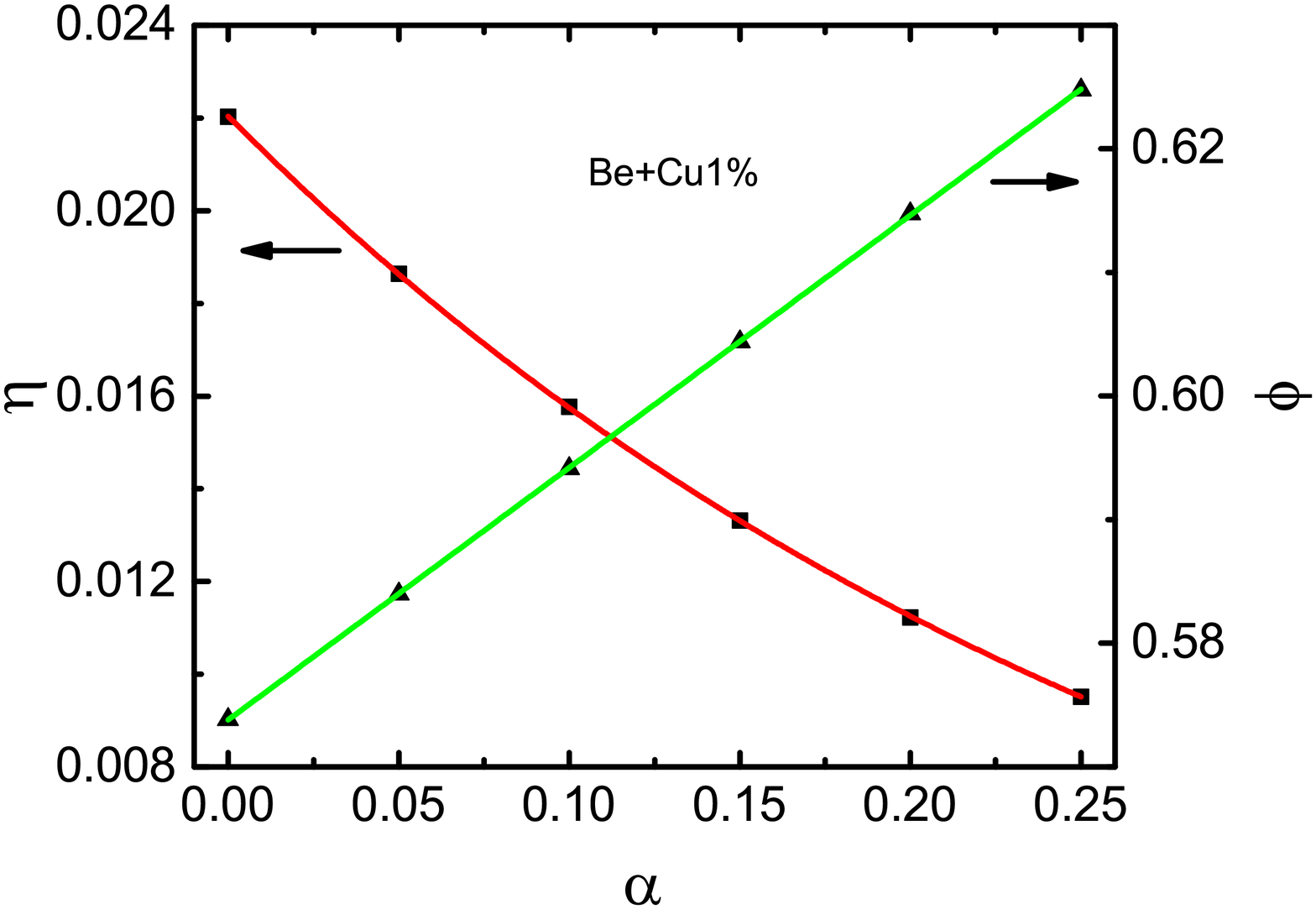}
      \caption{Variation of $\eta$ and $\phi$ with $\alpha$ in a Be+1\%Cu foil.\label{eta_phi_alpha_Be1} }
  \end{center}
\end{figure}

Now we discuss the results for Cu doped Be foil. Cu being a mid-Z element, its total extinction coefficient is much higher compared to low-Z ablator Be at higher frequencies as shown in Fig. \ref{KT-BeCu}. As the incident radiation consisting of low frequency Planckian spectrum and high frequency Gaussian spectrum passes through the foil, the hard x-rays get absorbed in the foil. This results in reduced preheat temperature at the rear surface of the foil. The results of shock velocity variation with radiation drive temperature are shown in Fig. \ref{TRvsUs_BeCu}. We note from this plot that the shock velocity reduces on adding 1\% Cu dopant. To explain this, we have plotted Rosseland mean opacity as a function of temperature at a density of 2 g/cc for pure Be and Cu doped Be in Fig. \ref{op-BeCu}. We note that doped ablator has higher opacity compared to pure Be that leads to the increased albedo. Consequently, the shock velocites should decrease in doped Be compared to pure Be. Moreover, we have shown the reduction in shock velocities at all temperatures for Be foils doped with 1\% Cu for $\alpha=0$ in Fig. \ref{TRvsUs_a0}. This variation is also observed for higher values of $\alpha$ as shown in Fig. \ref{TRvsUs_BeCu}. We also observe that shock velocities increase with $\alpha$ for all temperatures ranging from 120 eV to 200 eV. Thus, for a higher fraction of M-band energy densities, the shock velocities in a Be+1\%Cu foil approach those in pure Be. The behaviour is different from that of pure Be at higher temperatures. To explain this we have plotted the total extinction coefficient of Be ablator with 1\% Cu  as a function of frequency in Fig. \ref{KT-I-BeCu1-120eV} at a temperature of 120 eV and density of 2 g/cc. In the same graph, we have also shown the incident intensities for various $\alpha$ values at 120 eV. As in the case of pure Be, the total extinction coefficient of Be+1\% Cu is found to fall very sharply with frequency. At the frequency at which the Planckian spectrum peaks, the total extinction coefficient is as high as 796 $cm^2/g$, but falls to 21 $cm^2/g$ at the peak of the Gaussian spectrum. With increase in $\alpha$ at 120 eV, the effective opacity reduces significantly as in case of pure Be and hence the shock velocity increases. To explain the increase in $U_s$ with $\alpha$ at higher temperatures, we have plotted the total extinction coefficient ($K_t$) of Be+1\%Cu as a function of frequency in Fig. \ref{KT-I-BeCu1-200eV} at a temperature of 200 eV and density of 2 g/cc. In the same graph, we have also shown the incident intensities for various $\alpha$ values at 200 eV. We observe that $K_t$ corresponding to peak of Planckian reduces to a value of 227 $cm^2/g$ at higher temperature of 200 eV. This fact was also observed for pure Be at higher temperatures. The main difference in case of doped Be is the significant contribution of incident drive for various values of $\alpha$ in the frequency range of 1-1.5 keV due to the presence of extra lines in $K_t$ spectrum. As $\alpha$ increases, the effective opacity starts reducing, resulting in increase of shock velocity even at 200 eV. As for Be, we have also fitted $T_r$ vs $U_s$ for all values of $\alpha$ in accordance with Eq.\ref{TrUs}. A representative fit is shown in Fig. \ref{TRvsUs_BeCu1_fit} for $\alpha$ = 0.2. The values of $\eta(\alpha)$ and $\phi(\alpha)$ obtained for all $\alpha$ values are shown in Fig. \ref{eta_phi_alpha_Be1}. As previously done for Be, $\eta$ and $\phi$ are found to satisfy  the power law and linear relationship respectively, with $\alpha$. The fitting coefficients a, b, c and d are obtained as 0.02204, 0.03471, 0.5738 and 0.2041 respectively, for Be+1\% Cu.

\subsection{Shock breakout temperature scaling}
The shock front moves through an ablator foil with the shock velocity. As explained in the earlier section, the shock velocity increases with increase in $\alpha$ because of the decrease in effective opacity. The dependence of shock velocity on $\alpha$ is also found to be weak. As the shock front reaches the rear surface of the foil, shock breakout occurs and the material attains the shock breakout temperature ($T_s$). Due to rarefaction, the breakout temperature obtained at the rear surface is lower compared to the shock temperature within the ablator.   

However, the shock breakout temperature has a strong dependence on $\alpha$. As the energy density in the hard x-ray increases, more photons move at a higher speed causing more preheating. 
In this section, we relate the radiation temperature ($T_r$) to the shock breakout temperature ($T_s$) \cite{Drake}. Here, we note that the thermodynamic variables behind and ahead of the shock front can be measured either in the laboratory or the shock frame of reference. In the laboratory frame of reference, the upstream fluid into which the shock enters is at rest. The downstream fluid is however disturbed by the shock and the medium acquires a particle velocity. The shock frame is the frame of reference in which the shock is at rest. The upstream fluid into which the shock moves is designated by the subscript 1 and the downstream fluid is denoted by the subscript 2. The rate at which the upstream material and the shock approach one another is called the shock velocity.
For strong shocks,
\begin{equation}
P_2 \sim \rho_1 U_s^2.
\end{equation}
where $P_2$ is the pressure in the downstream fluid , $\rho_1$ is the density in upstream fluid and $U_s$ is the shock velocity. 
Similarly, for temperature we have,
\begin{equation}
P_2=\frac{(Z_2+1)k_BT_2\rho _2}{Am_p},
\end{equation}
where $Z_2$, $T_2$ and $\rho_2$ are the charge state, temperature and density in the downstream fluid. A and $m_p$ are the mass number and proton mass, respectively. 
Thus, we get the relation between $U_s$ and $T_s$ as
\begin{equation}
U_s=k\sqrt{T_s}.
\end{equation}
Substituting in Eq. \ref{TrUs}, we get
\begin{equation}
T_r=\eta_s T_s^{\phi_s}.
\end{equation}

Similarly, in case of a non Planckian spectrum, the shock breakout temperature varies with both the radiation temperature and $\alpha$ as
\begin{equation}\label{TrTs}
T_r=\eta_s(\alpha)T_s^{\phi_s(\alpha)}.
\end{equation}

\begin{figure}    
   \begin{center}
       \includegraphics[width=0.8\linewidth]{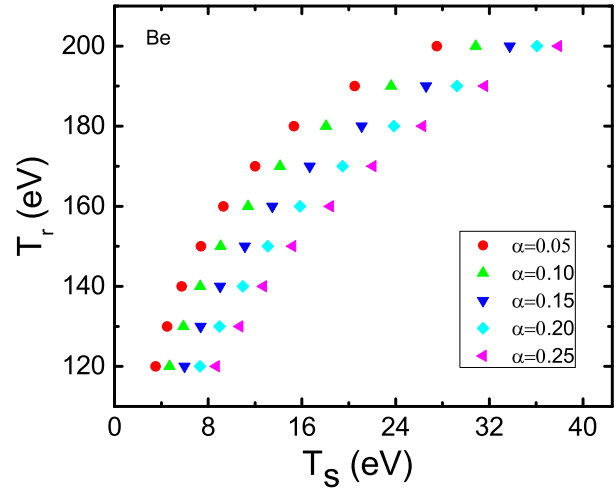}
      \caption{Variation of $T_r$ with $T_{s}$ in a Be foil for values of $\alpha$ varying from 0.05 to 0.25.\label{TRvsTs_Be} }
  \end{center}
\end{figure}

\begin{figure}    
   \begin{center}
       \includegraphics[width=0.8\linewidth]{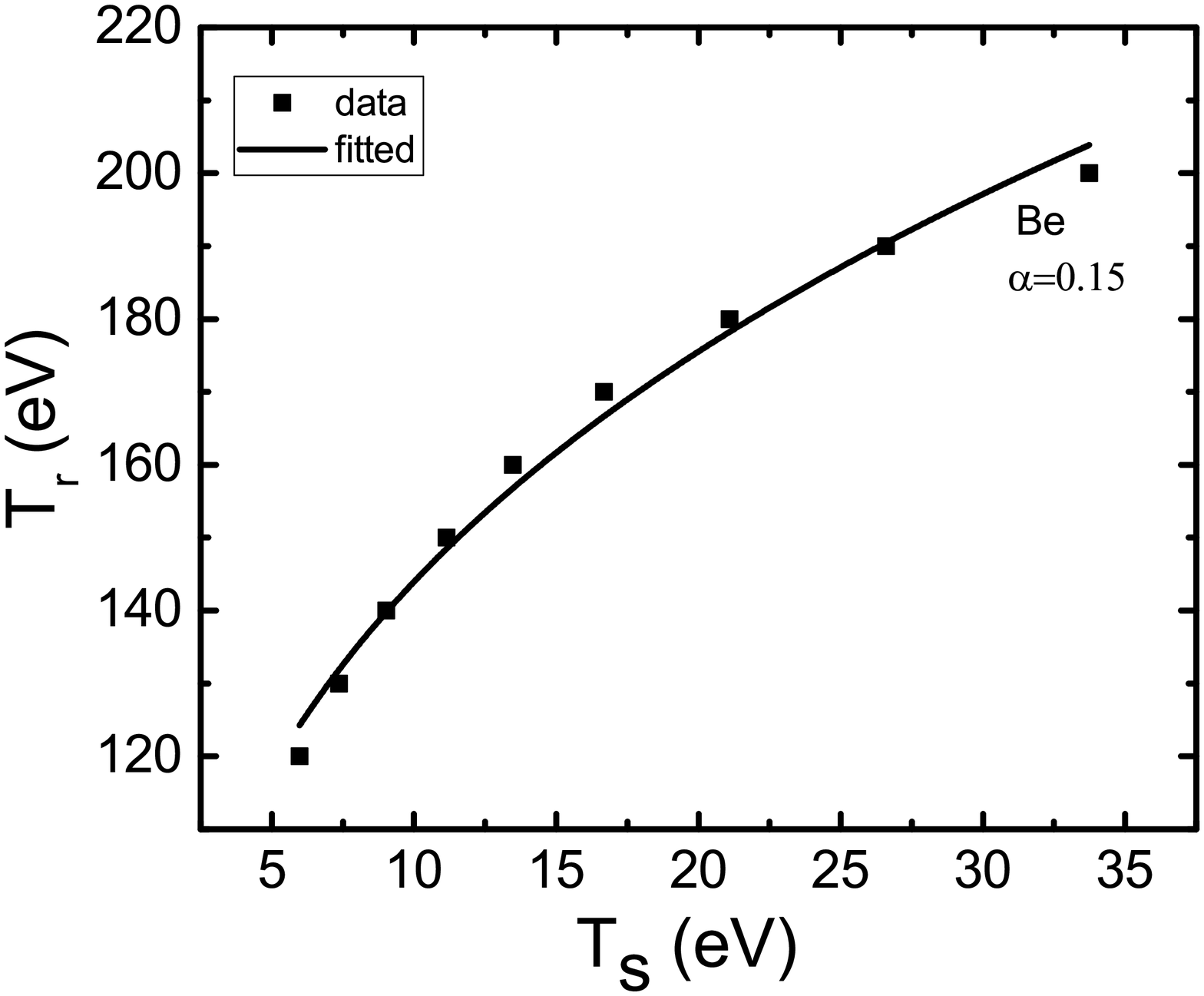}
      \caption{Fitting of $T_r$ with $T_s$ in a Be foil for $\alpha$ = 0.15.\label{TRvsTs_Be_fit}  }
  \end{center}
\end{figure}

\begin{figure}    
   \begin{center}
       \includegraphics[width=0.8\linewidth]{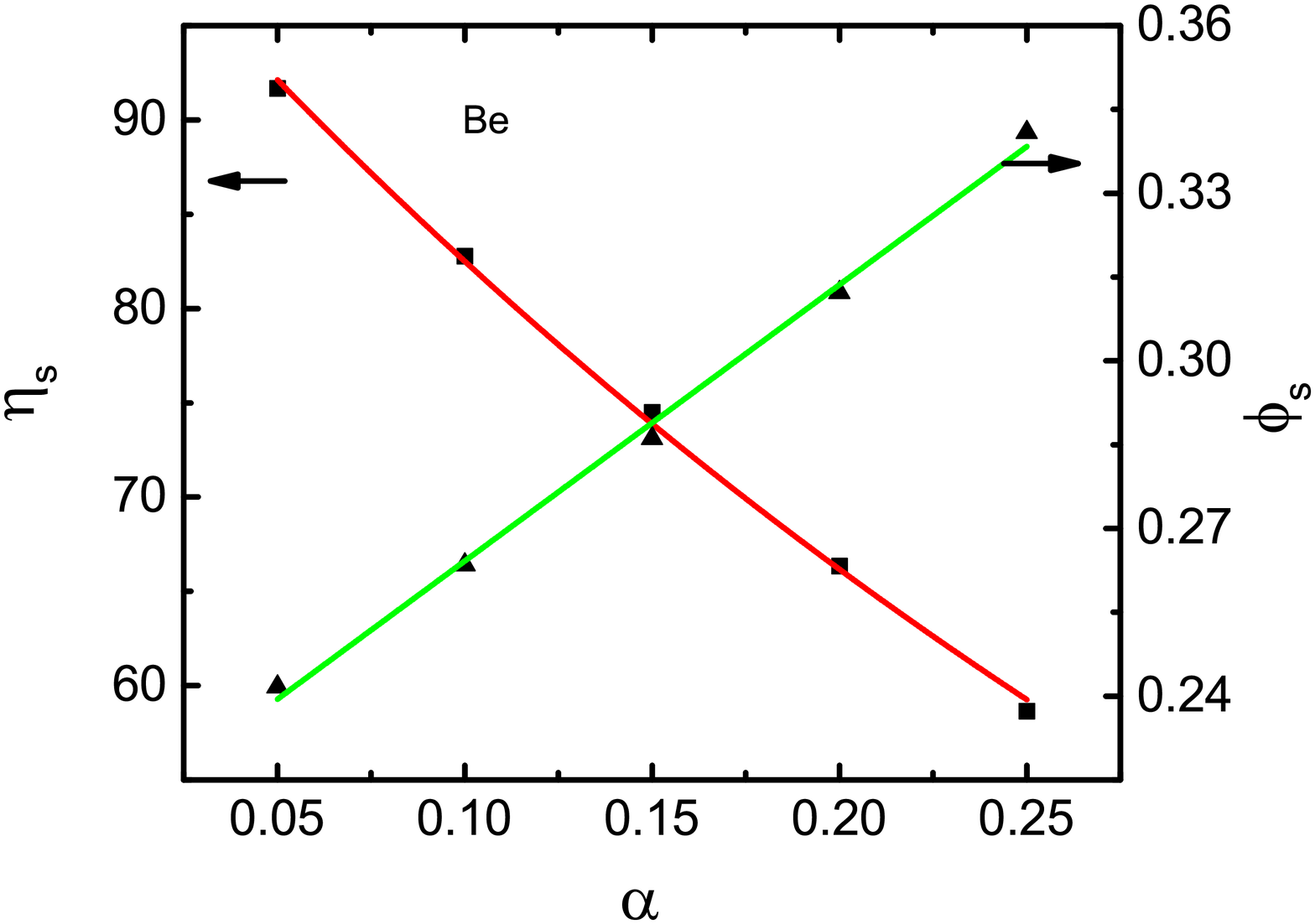}
      \caption{Variation of $\eta_s$ and $\phi_s$ with $\alpha$ in a Be foil.\label{par-Tr-Ts_alpha_Be}}
  \end{center}
\end{figure}

The simulation results of shock breakout temperature ($T_s$) are shown in Fig. \ref{TRvsTs_Be} for pure Be foil driven by various drive temperatures ($T_r$) in the range of 120 to 200 eV. We have also shown the effect of variation of $\alpha$ in the same plot. We observe significant variation in shock breakout temperature with $T_r$ as expected. Also, the shock breakout temperatures are found to increase with $\alpha$ as predicted by Eq. \ref{TrTs}. In Fig. \ref{TRvsTs_Be_fit}, radiation temperature has been fitted as a function of shock breakout temperature for $\alpha=0.15$. The fitting coefficients have been shown in Fig. \ref{par-Tr-Ts_alpha_Be}. As for the shock velocity scaling, $\eta_s$ is found to have a power law dependence on $\alpha$, viz. $\eta_s(\alpha)=a_s.b_s^\alpha$. Similarly, $\phi_s$ is found to vary linearly with $\alpha$ as $\phi_s(\alpha)=c_s+d_s.\alpha$. The values of $a_s,b_s,c_s$ and $d_s$ are 102.9, 0.1101, 0.2147 and 0.4956, respectively for Be.

\begin{figure}    
   \begin{center}
       \includegraphics[width=0.8\linewidth]{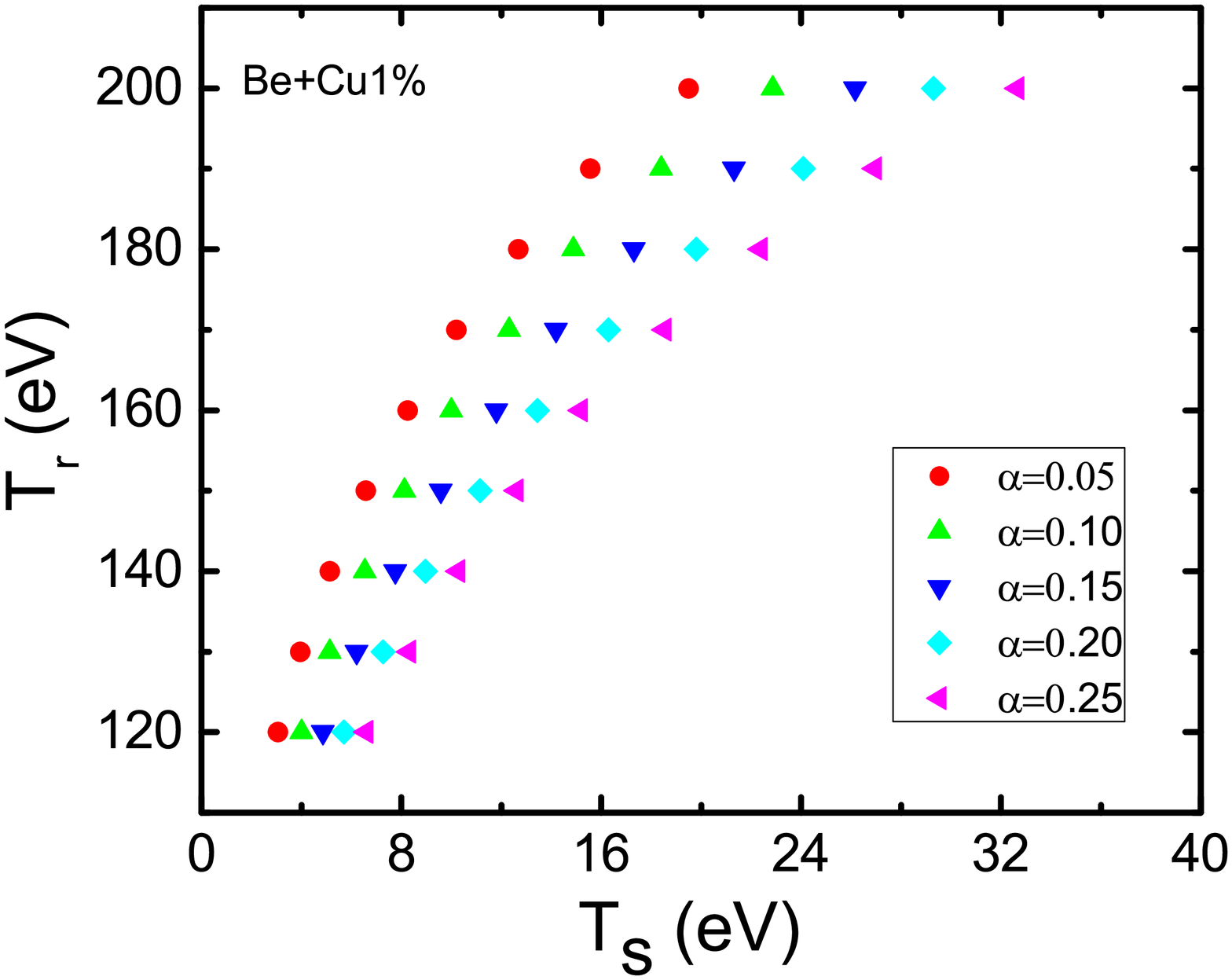}
      \caption{Variation of $T_r$ with $T_{s}$ in a Be+1\%Cu foil for values of $\alpha$ varying from 0.05 to 0.25.\label{TRvsTs_BeCu1}}
  \end{center}
\end{figure}

\begin{figure}    
   \begin{center}
       \includegraphics[width=0.8\linewidth]{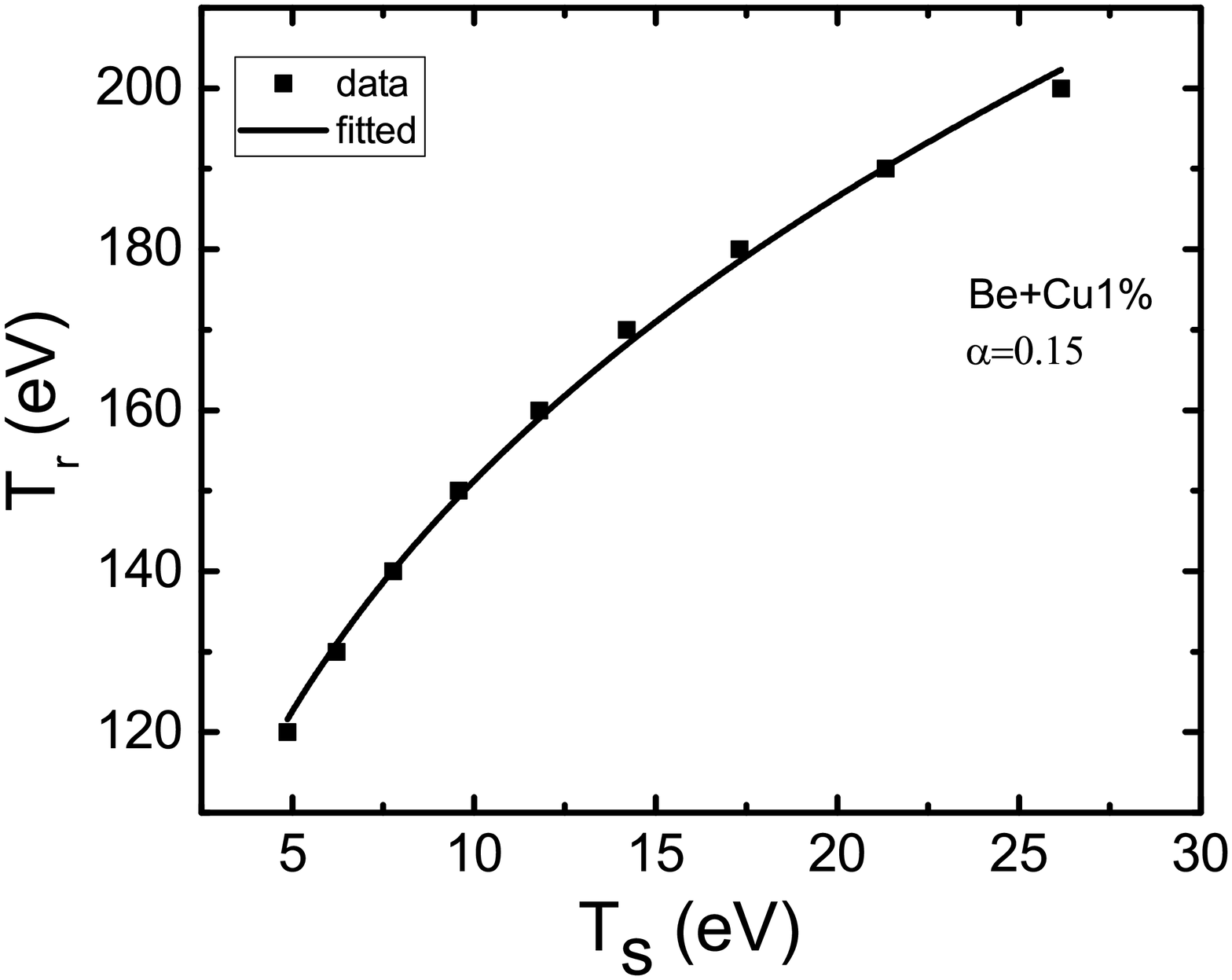}
      \caption{Fitting of $T_r$ with $T_s$ in a Be+1\%Cu foil for $\alpha$ = 0.15.\label{TRvsTs_BeCu1_fit}  }
  \end{center}
\end{figure}

\begin{figure}    
   \begin{center}
       \includegraphics[width=0.8\linewidth]{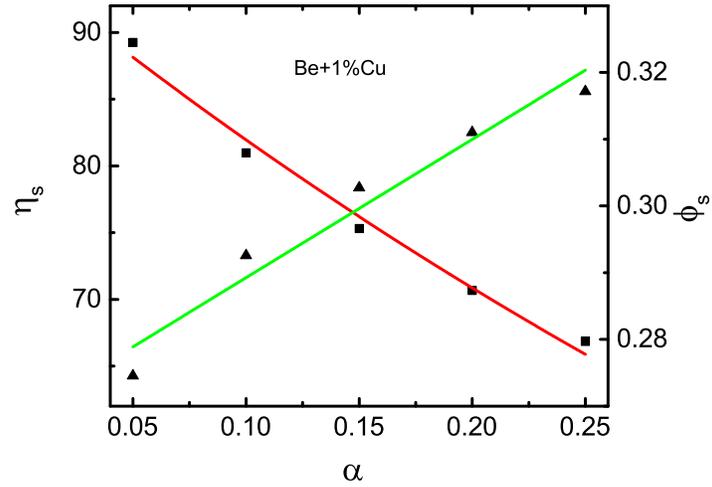}
      \caption{Variation of $\eta_s$ and $\phi_s$ with $\alpha$ in a Be+1\%Cu foil.\label{par-Tr-Ts_alpha_BeCu1}}
  \end{center}
\end{figure}

On doping Be with Cu, the shock breakout temperature is found to reduce considerably at all temperatures as shown in Fig. \ref{TRvsTs_BeCu1} for Be+1\% Cu foil. The data for drive temperature is fitted against shock breakout temperature for all alpha values by using Eq. \ref{TrTs}  and the result is shown in Fig. \ref{TRvsTs_BeCu1_fit} for a representative case of $\alpha$=0.15. The fitting coefficients $\eta_s $ and $\phi_s$ for Be+1\%Cu are shown in Fig. \ref{par-Tr-Ts_alpha_BeCu1}. Taking the dependence of $\eta_s$ and $\phi_s$ on $\alpha$ as power law and linear, the values of coefficients $a_s,b_s,c_s$ and $d_s$ are obtained as 94.8, 0.2333, 0.2685 and 0.2073, respectively. 

\subsection{Preheat temperature scaling}
The high energy photons ($h\nu \geq 2 keV $) in a non-Planckian radiation drive transfer energy ahead of the shock front in an ablator thus causing preheating. By considering the additional energy flux carried away from the surface of the front by radiation, the maximum preheat temperature ($T_{ph}$) ahead of the shock front is found to be proportional to the radiation flux (S) emerging from the shock front $S=\sigma T_s^4$ \cite{Zeldovich}. Thus, in case of a subcritical shock, the radiant energy ($\epsilon$) absorbed in the preheating region is used up only in raising the gas temperature. Hence, $\epsilon=\alpha T_{ph}^\beta$ and $-S=\epsilon\rho_0 U_s$ at the shock front. Writing $U_s$ and $T_s$ in terms of $T_r$, we obtain the following scaling relation between $T_{ph}$ and $T_r$
\begin{equation}\label{TrTph}
T_r=\eta_{ph}(\alpha)T_{ph}^{\phi_{ph}(\alpha)}.
\end{equation} 

\begin{figure}    
   \begin{center}
       \includegraphics[width=0.8\linewidth]{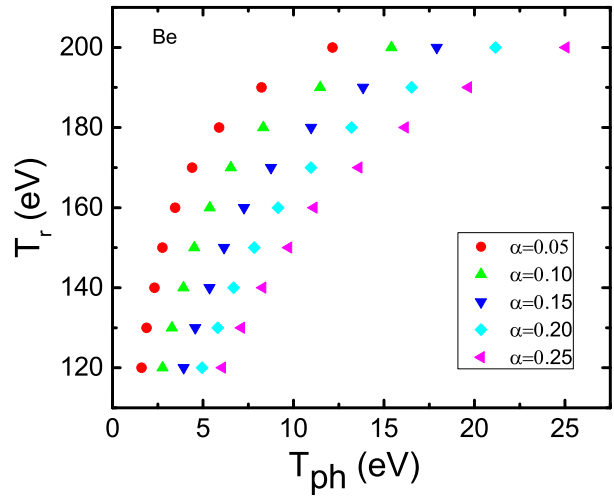}
      \caption{Variation of $T_r$ with $T_{ph}$ in a Be foil for values of $\alpha$ varying from 0.05 to 0.25.\label{TRvsTph_Be} }
  \end{center}
\end{figure}

\begin{figure}    
   \begin{center}
       \includegraphics[width=0.8\linewidth]{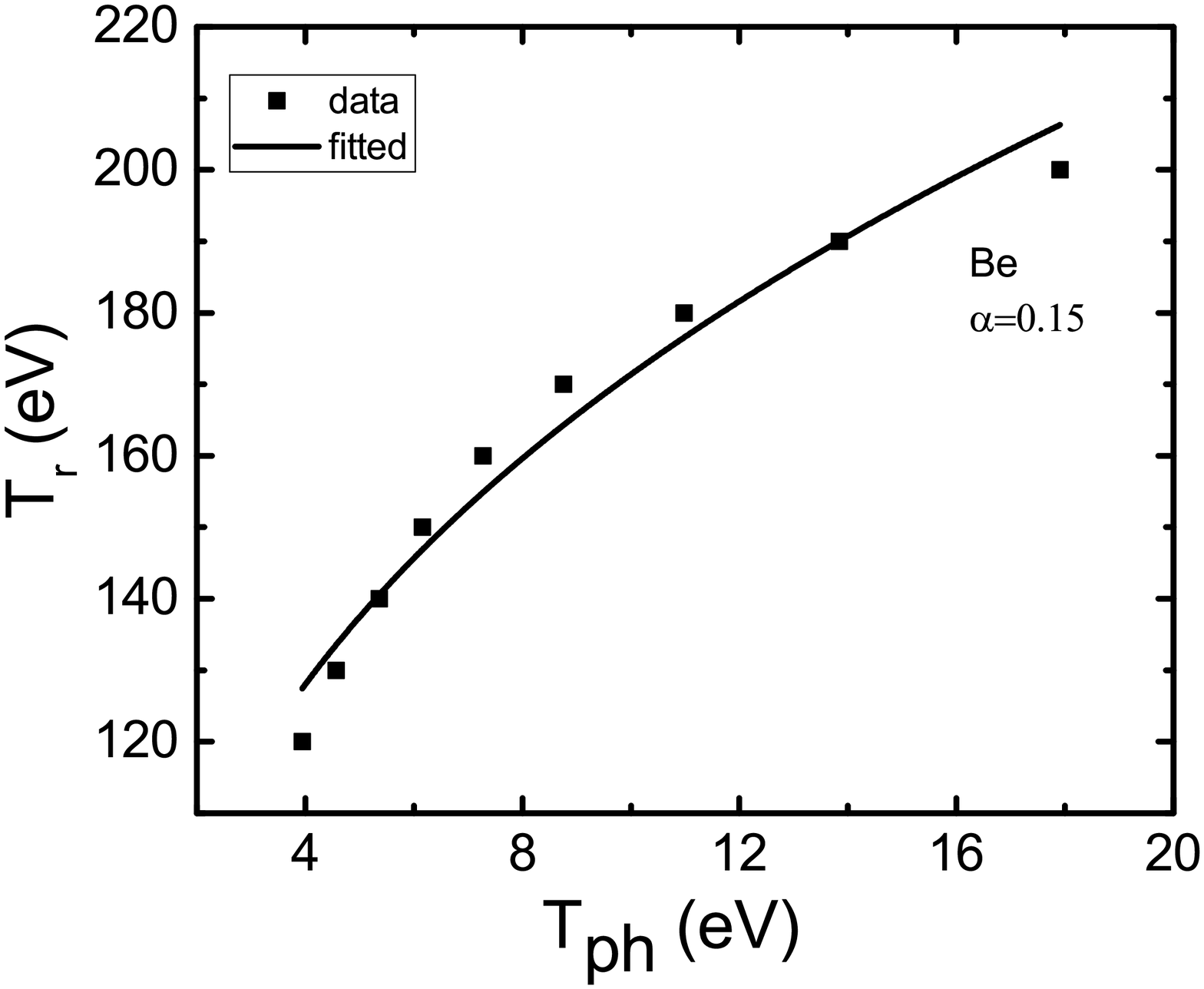}
      \caption{Fitting of $T_r$ with $T_{ph}$ in a Be foil for $\alpha$ = 0.15.\label{TRvsTph_Be_fit}  }
  \end{center}
\end{figure}

\begin{figure}    
   \begin{center}
       \includegraphics[width=0.8\linewidth]{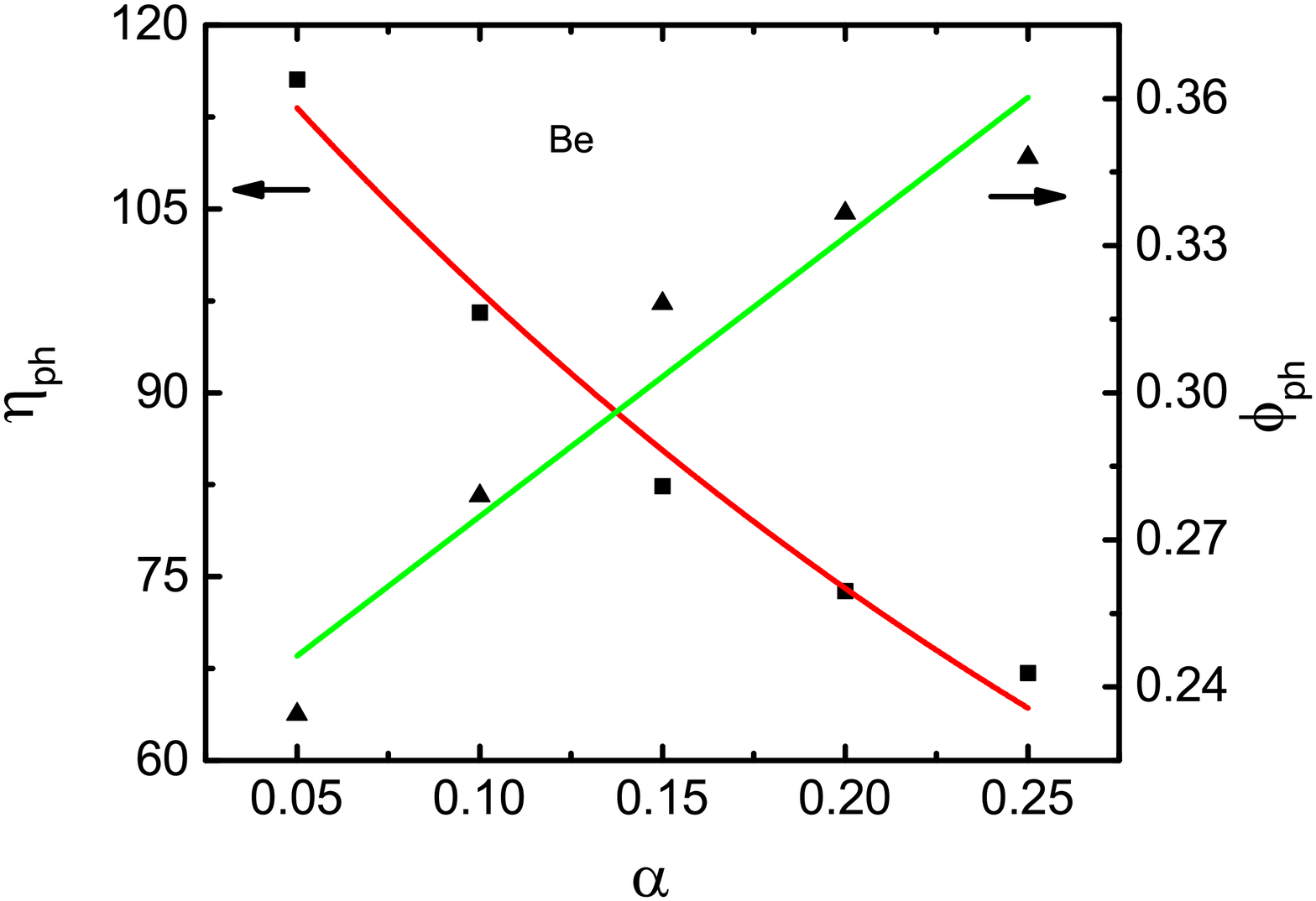}
      \caption{Variation of $\eta_{ph}$ and $\phi_{ph}$ with $\alpha$ in a Be foil.\label{par-Tr-Tph_alpha_Be}}
  \end{center}
\end{figure}

In Fig. \ref{TRvsTph_Be}, the variation in maximum preheat temperatures has been plotted for various incident radiation temperatures from 120 eV to 200 eV. Effect of changing $\alpha$ from 0.05 to 0.25 is also shown in the same plot. Similar to shock breakout temperature, we observe significant variation in maximum preheat temperature with $T_r$ as expected. As the values of $T_s$ are very low for $\alpha=$0, we have presented the results from $\alpha=$0.05 to $\alpha=$0.25. Increase in the fraction of hard x-ray energy density leads to higher preheating as expected. The data for drive temperature is fitted against maximum preheat temperature for $\alpha=$0.05 to $\alpha=$0 by using Eq. \ref{TrTph} and the result is shown in Fig. \ref{TRvsTph_Be_fit} for a representative case of $\alpha=$0.15.  Also, similar to the shock velocity and  shock breakout temperature scaling, $\eta_{ph}$ and $\phi_{ph}$ are found to have a power law dependence on $\alpha$, viz. $\eta_{ph}(\alpha)=a_{ph}.b_{ph}^\alpha$ and $\phi_{ph}(\alpha)=c_{ph}+d_{ph}.\alpha$. The variation of fitting coefficients $\eta_{ph}$ and $\phi_{ph}$ is shown in Fig. \ref{par-Tr-Tph_alpha_Be} as a function of $\alpha$. The values of $a_{ph},b_{ph},c_{ph}$ and $d_{ph}$ are obtained as 130.5, 0.05894, 0.2178 and 0.5695 respectively for Be.

\begin{figure}    
   \begin{center}
       \includegraphics[width=0.8\linewidth]{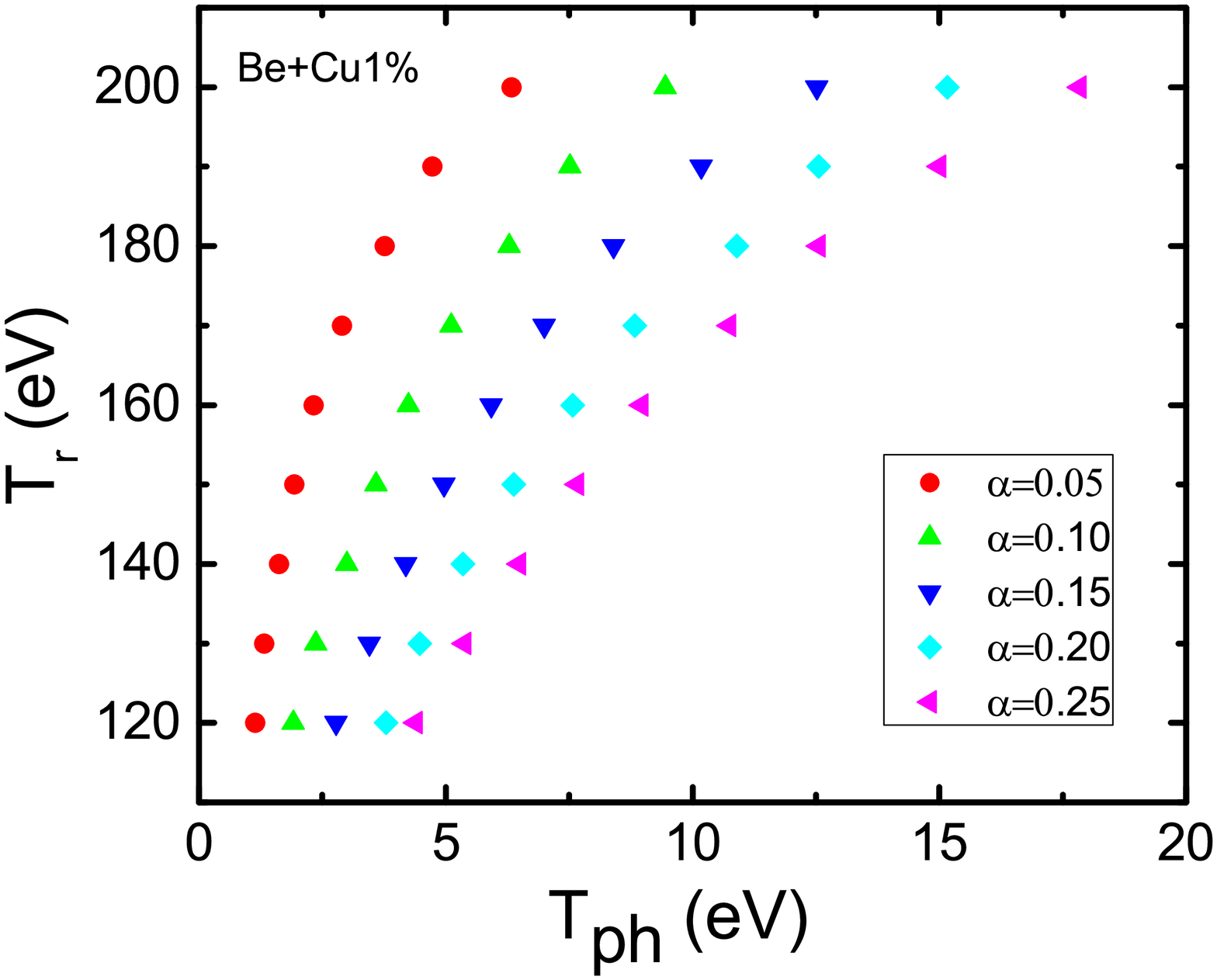}
      \caption{Variation of $T_r$ with $T_{ph}$ in a Be+1\%Cu foil for values of $\alpha$ varying from 0.05 to 0.25.\label{TRvsTph_BeCu1} }
  \end{center}
\end{figure}

\begin{figure}    
   \begin{center}
       \includegraphics[width=0.8\linewidth]{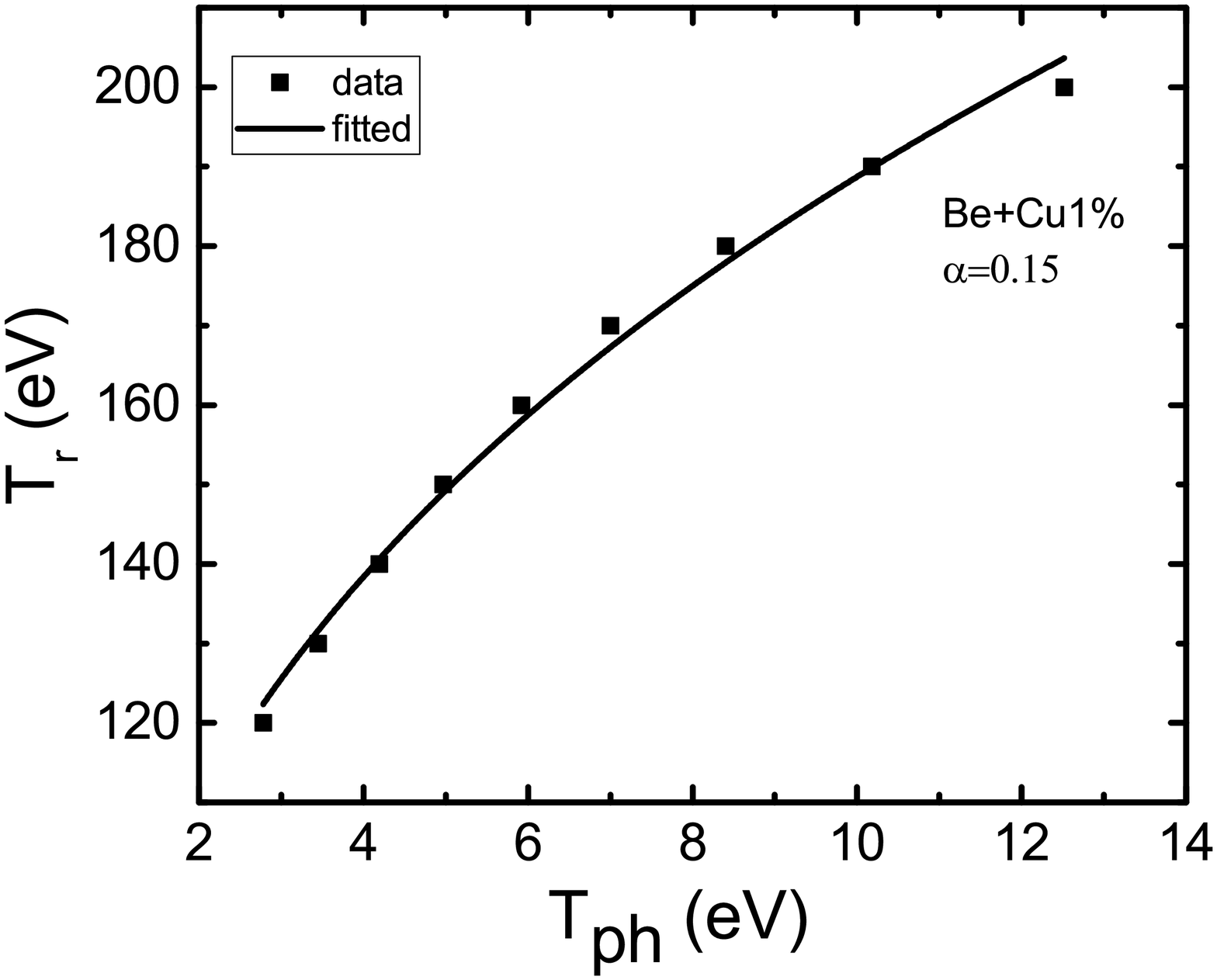}
      \caption{Fitting of $T_r$ with $T_{ph}$ in a Be+1\%Cu foil for $\alpha$ = 0.15.\label{TRvsTph_BeCu1_fit}  }
  \end{center}
\end{figure}

\begin{figure}    
   \begin{center}
       \includegraphics[width=0.8\linewidth]{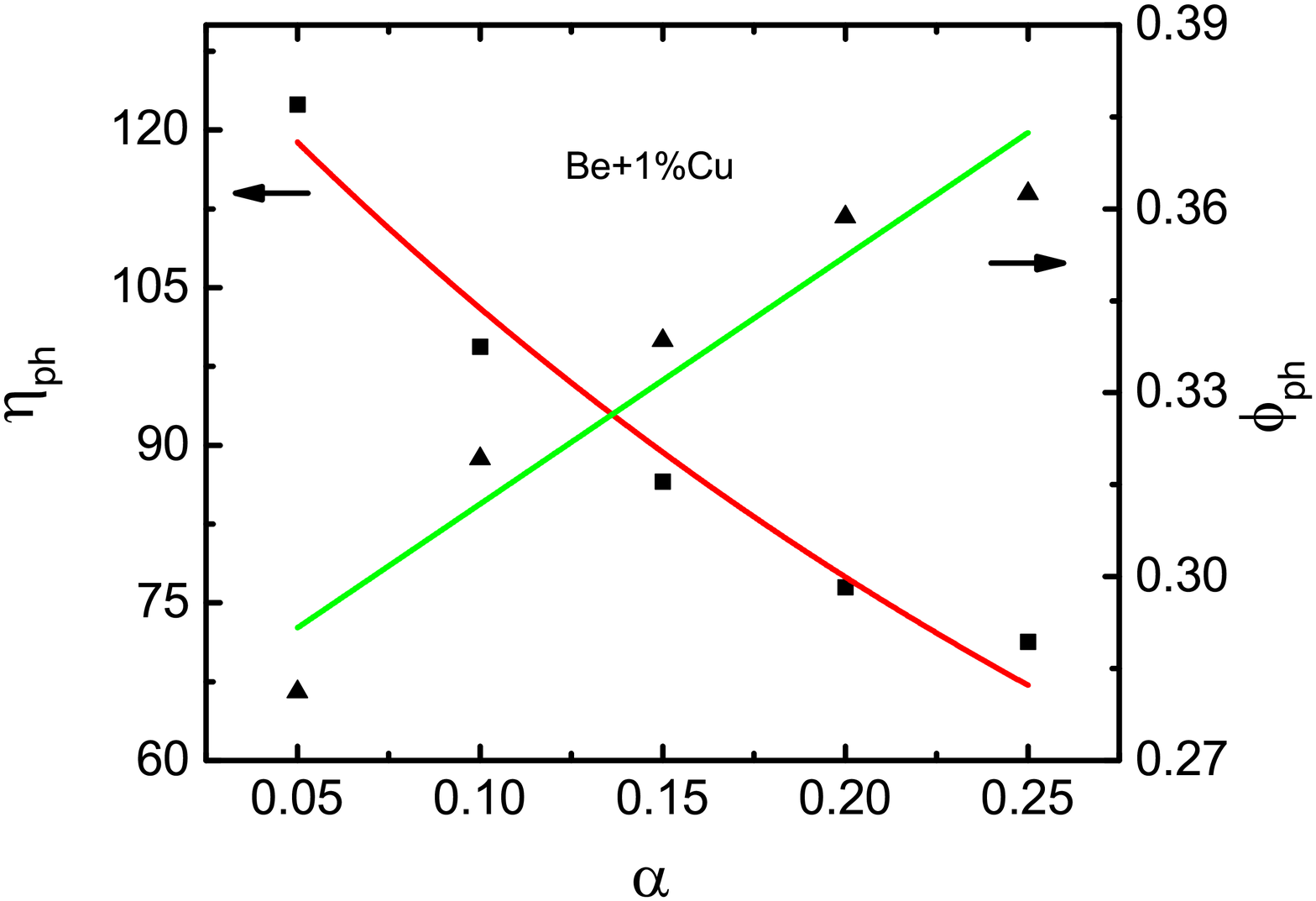}
      \caption{Variation of $\eta_{ph}$ and $\phi_{ph}$ with $\alpha$ in a Be+1\%Cu foil.\label{par-Tr-Tph_alpha_BeCu1}}
  \end{center}
\end{figure}

On doping Be with Cu, the maximum preheat temperature is found to reduce considerably at all temperatures as shown in Fig. \ref{TRvsTph_BeCu1} for Be+1\% Cu foil. The data for drive temperature is fitted against maximum preheat temperature for all $\alpha$ values by using Eq. \ref{TrTph} and the result is shown in Fig. \ref{TRvsTph_BeCu1_fit} for a representative case of $\alpha=$0.15. The fitting coefficients $\eta_{ph}$ and $\phi_{ph}$ for Be+1\%Cu are plotted as a function of $\alpha$ in Fig. \ref{par-Tr-Tph_alpha_BeCu1}. Taking the dependence of $\eta_{ph}$ and $\phi_{ph}$ as power law and linear on $\alpha$, the values of coefficients $a_{ph},b_{ph},c_{ph}$ and $d_{ph}$ are obtained as 137.1, 0.05763, 0.2714 and 0.4040 respectively.

\begin{figure}    
   \begin{center}
       \includegraphics[width=0.8\linewidth]{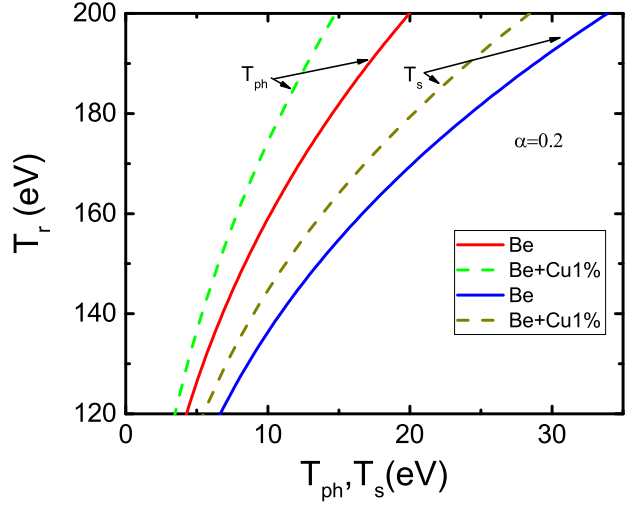}
      \caption{Variation of $T_{ph}$ and $T_s$ with $T_r$ in case of Be and Be+1\%Cu ablator foils for $\alpha$ = 0.2.\label{TRvsTph_Ts_compare} }
  \end{center}
\end{figure}

In Fig. \ref{TRvsTph_Ts_compare}, we compare the preheat and shock breakout temperatures in pure and doped Be ablators. Significant reduction in preheat and shock breakout temperatures are observed on doping.

By measuring both $T_s$ and $T_{ph}$ in an ablator foil (either Be or Be +1\%Cu) of thickness 80 $\mu m$, and simultaneously solving the scaling relations viz., Eqs. \ref{TrTs} and  \ref{TrTph}, we can obtain the values of $T_r$ and $\alpha$ for an unknown constant temperature drive.

\subsection{Mass ablation rate scaling}
\begin{figure}    
   \begin{center}
       \includegraphics[width=0.8\linewidth]{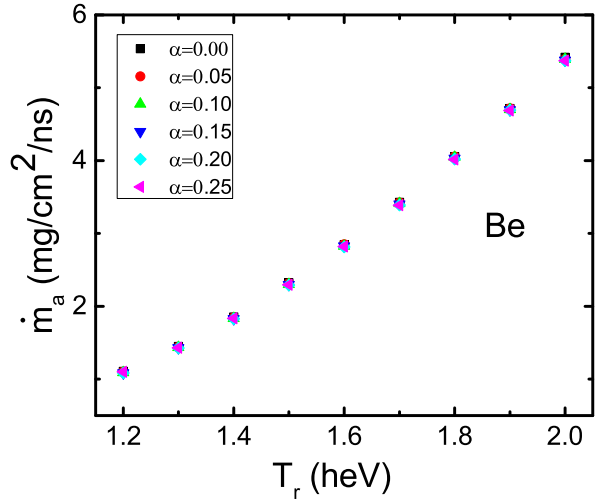}
      \caption{Variation in mass ablation rate $\dot{m_a}$ with $T_r$ in a Be foil for values of $\alpha$ varying from 0 to 0.25.\label{marTr_Be} }
  \end{center}
\end{figure}

\begin{figure}    
   \begin{center}
       \includegraphics[width=0.8\linewidth]{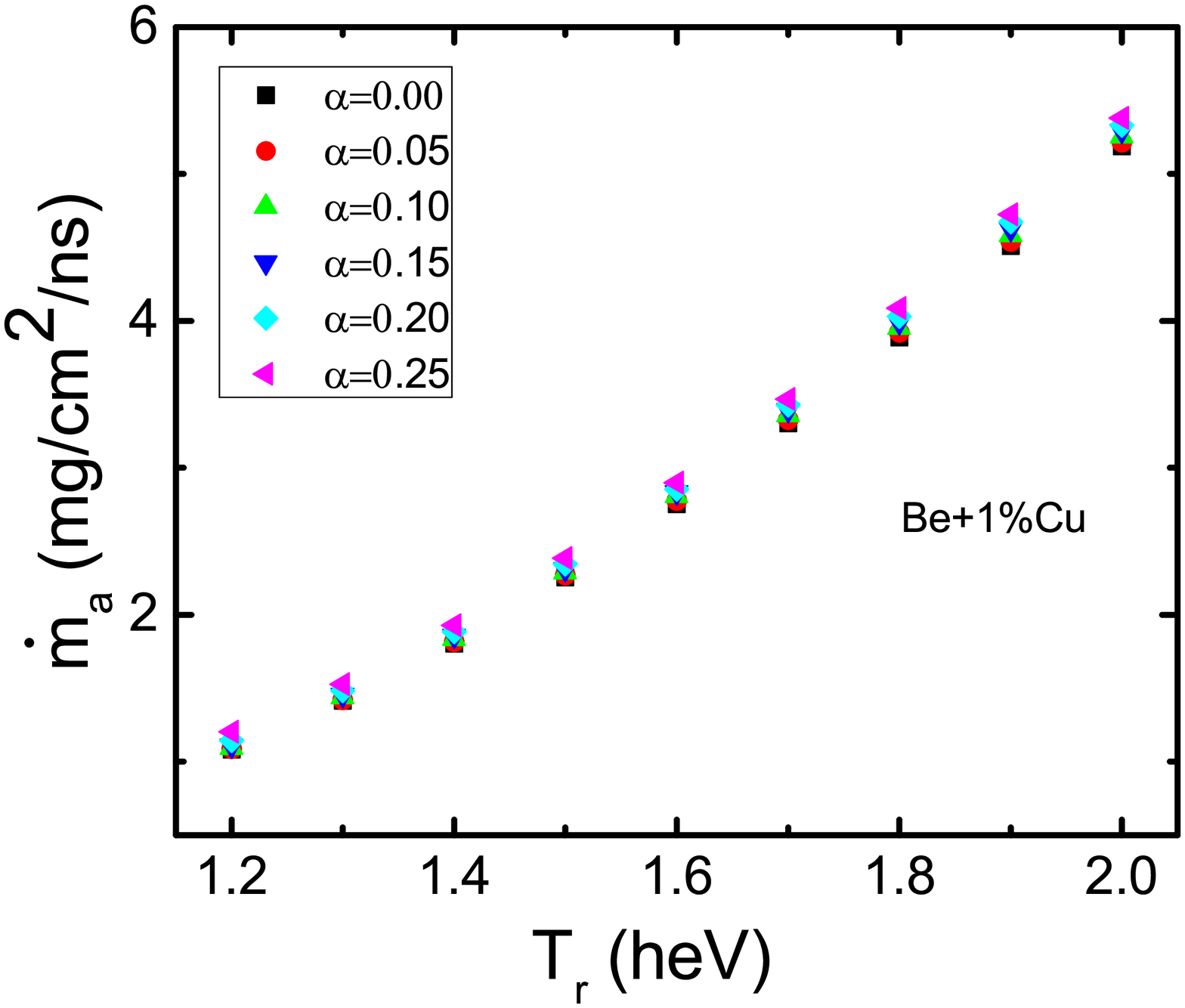}
      \caption{Variation in mass ablation rate $\dot{m_a}$ with $T_r$ in a Be+1\%Cu foil for values of $\alpha$ varying from 0 to 0.25.\label{marTr_BeCu1} }
  \end{center}
\end{figure}

\begin{figure}    
   \begin{center}
       \includegraphics[width=0.8\linewidth]{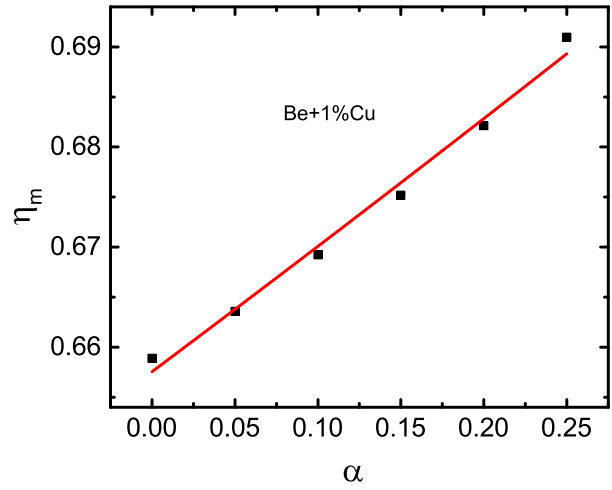}
      \caption{Variation of $\eta_m$ with $\alpha$ in a Be+1\%Cu foil for values of $\alpha$ varying from 0 to 0.25.\label{mar_eta_BeCu1_fit} }
  \end{center}
\end{figure}

\begin{figure}    
   \begin{center}
       \includegraphics[width=0.8\linewidth]{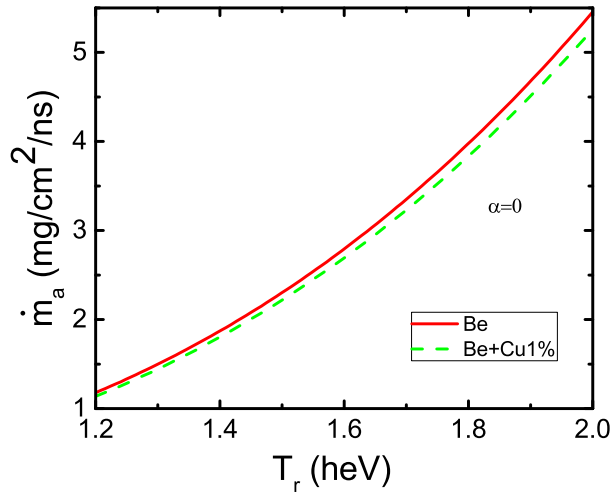}
      \caption{Comparison of mass ablation rates in pure Be and Be+1\%Cu foils for various values of $T_r$. Here, $\alpha$ is equal to 0.\label{mar_Be_BeCu1}}
  \end{center}
\end{figure}

In this subsection, we compare the performance of Be and doped Be ablators in terms of mass ablation rate. For laser or radiation driven ablation phenomena, the mass of material ablated outwards is closely related to energy transport inwards \cite{Lindl1995,Saillard}. The mass ablation rate is the most commonly measured experimental quantity. X-ray burnthrough flux is obtained at the rear side of the ablator sample using a streaked x-ray imager (SXI) in a hohlraum experiment. By noting the time of arrival of soft x-ray photons, planar ablation rate is obtained from a knowledge of the ablator density and thickness \cite{POP18Olson}. It is a measure of the depth of material penetrated by the heat front during the drive pulse. The mass ablation rate ($\dot{m_a}$) and ablation pressure ($P_a$) have been extensively scaled with laser intensity, laser wavelength and target atomic number \cite{Dahmani1,Dahmani2}. They have also been scaled with the radiation temperature in x-ray driven ablation by taking the spectral form of incident drive as complete Planckian \cite{POP18Olson,Kline}. In this section, we compare the performance of Be and doped Be ablators driven by constant temperature drive in terms of $\dot{m_a}$. We have also obtained the scaling relations for mass ablation rate as a function of drive temperature and fraction of hard x-ray energy density in the incident drive spectrum. As explained in section 3.1, ablation pressure $P \sim \dot{m_a}U_p \sim (1-\alpha_d)T_r^4/U_p$, that leads to $\dot{m_a} \sim (1-\alpha_d)T_r^4/U_p^2$ \cite{MishraHEDP2018,Hatchett}. Due to low opacity of the heated ablator and taking the speed of exhaust $U_p \sim \sqrt{T_r}$, we expect the mass ablation rate to roughly scale as $T_r^3$. We note that both  experiments and simulations show good agreement with the expected $T_r^3$ dependence of $\dot{m_a}$ \cite{POP18Olson}. In Fig. \ref{marTr_Be}, the simulated mass ablation rates have been plotted as a function of radiation temperatures for various $\alpha$ values in Be foil. As more energy is deposited for higher incident temperatures, mass ablation rate keeps increasing with temperature. However, the values of $\dot{m_a}$ do not vary with $\alpha$ and is found to vary with $T_R$ as $\dot{m_a}=0.6820T_r^3$. This shows that the presence of hard x-rays in the source do not affect the ablation properties in pure Be ablator. Mass ablation rates have been plotted for constant radiation temperatures (120-200eV) in a Be+1\%Cu foil for $\alpha$ values ranging from 0 to 0.25. $\dot{m_a}$ is found to increase slightly with $\alpha$ viz. $\dot{m_a}=\eta_m(\alpha)T_r^3$ as shown in Fig. \ref{marTr_BeCu1}. Here, we have fitted $\eta_m(\alpha)=a_m b_m^\alpha $ with $a_m=0.6576$ and $b_m=1.208$ as shown in Fig. \ref{mar_eta_BeCu1_fit}. Further, we have also compared $\dot{m_a}$ for pure Be and doped Be in case of pure Planckian spectrum with temperature in Fig. \ref{mar_Be_BeCu1}. We find that doped Be has slightly less $\dot{m_a}$ compared to pure Be in this case. It may be further noted that both pure and doped Be follow $T_r^3$ law but doped Be shows slow dependence on $\alpha$ for a given temperature. This is the result of higher total extinction coefficient of Be+1\%Cu compared to pure Be at higher photon energies. Hence, as the fraction of M-band energy density increases, in doped Be,  the energy absorption in the hard x-ray spectrum also increases leading to slight increase in $\dot{m_a}$ with $\alpha$. In case of pure Be, the extinction coefficient being small at higher photon energies, $\dot{m_a}$ is independent of $\alpha$ for a given temperature.  

\section{Conclusion}\label{conc}
In this paper, we have quantitatively analyzed the effect of soft and hard x-rays on doped and undoped Be ablators using radiation hydrodynamics simulations. Accurate scaling relations have been obtained for shock velocity, shock breakout temperature, maximum preheat temperature and mass ablation rate with drive temperature and fraction of hard x-ray energy density in planar foils of the ablators. Effect of hard x-rays is found to be significant on the shock and preheat temperatures. Shock velocities are found to slightly increase with hard x-ray fraction in pure Be upto 170 eV incident radiation temperature. Above this temperature, the shock velocities do not change with $\alpha$. Slight increase in shock velocity is observed for Be doped with 1\% Cu on increasing $\alpha$. Further, significant reduction in both preheat and shock breakout temperature is observed on doping Be with 1\% Cu. Mass ablation rate, on the other hand remains nearly unaffected on doping. Thus, our study confirms that Be+1\% Cu ablator is superior to undoped Be as it prevents preheating without affecting the ablation properties. 
\label{concl}

\begin{thebibliography}{10}
\expandafter\ifx\csname url\endcsname\relax
  \def\url#1{\texttt{#1}}\fi
\expandafter\ifx\csname urlprefix\endcsname\relax\def\urlprefix{URL }\fi
\expandafter\ifx\csname href\endcsname\relax
  \def\href#1#2{#2} \def\path#1{#1}\fi
  
\bibitem{Atzeni}
S. Atzeni, J. Meyer-ter-Vehn, The Physics of Inertial Fusion (Oxford Science, Oxford, 2004).  

\bibitem{Lindl2004}
J.~D. Lindl, et \emph{al.}, The physics basis for ignition using indirect-drive
  targets on the {N}ational {I}gnition {F}acility, Phys. Plasmas 11 (2004) 339. 

\bibitem{olson2004}
R. E. Olson, et \emph{al.}, Shock propagation, preheat, and x-ray burnthrough in indirect-drive inertial confinement fusion ablator materials, Phys. Plasmas 11 (2004) 2778.

\bibitem{Robey}H. F. Robey, et \emph{al.}, Experimental measurement of Au M-band flux in indirectly driven double-shell implosions, Phys. Plasmas 12 (2005) 072701.

\bibitem{Troussel} Ph. Troussel et. \emph{al.}, Absolute radiant power measurement for the
Au M lines of laser-plasma using a calibrated
broadband soft X-ray spectrometer with flat-
spectral response, Rev. Sci. Instr. 85 (2014) 013503.

\bibitem{LiHEDP} L. Li, Measuring impact of M-band and soft X-rays on radiation-driven ablation
performance, High Energy Density Physics 25 (2017) 1. 

\bibitem{POP21}W. Y. Huo et. \emph{al.}, The radiation temperature and M-band fraction inside hohlraum on the SGIII-prototype
laser facility, Phys. Plasmas 21 (2014) 022704. 

\bibitem{POP18}R. P. J. Town et. \emph{al.}, Analysis of the National Ignition Facility
ignition hohlraum energetics experiments,  Phys. Plasmas 18 (2011) 056302. 

\bibitem{POP18AlTi}Y. Li et \emph{al.}, A novel method for determining the
M-band fraction in laser-driven gold hohlraums,  Phys. Plasmas 18  (2011) 022701. 

\bibitem{POP20Al}C. Zhang, et \emph{al.}, Investigation of radiation flux in certain band via the preheat of aluminum sample, Phys. Plasmas 20 (2013) 122706.

\bibitem{JAP58}L. Dasilva et. \emph{al.}, Simulations of temperature measurements of shock-heated solids, Journal App. Phys. 58 (1985) 3634.

\bibitem{OC53}A. Ng, Measurement of shock heating in laser-irradiated solids,  Optics Communications 53 (1985) 389.

\bibitem{preheat-Olson-2003} 
R. E. Olson, et. \emph{al.}, Preheat effects on shock propagation in indirect-drive Inertial Confinement Fusion ablator materials, Phys. Rev. Lett. 91 (2003) 235002.

\bibitem{POP26-2019}  
C. Zhang, et. \emph{al.}, Study of M-band X-ray preheating effect on shock propagation via streaked optical pyrometer system at SG-III prototype lasers, Phys. Plasmas 26 (2019) 012708. 

\bibitem{POP23-Cheng}B. Cheng et. \emph{al.}, Effects of preheat and mix on the fuel adiabat of an imploding capsule, Phys. Plasmas 23 (2016) 120702.

\bibitem{POP22thin}L. Ling et. \emph{al.}, A method for evaluation the mean preheat temperature in x-ray driven ablation, Phys. Plasmas 22 (2015) 032705.

\bibitem{POP18Olson}R. E. Olson et. \emph{al.}, X-ray ablation rates in inertial confinement fusion capsule materials, Phys. Plasmas 18, 032706 (2011).

\bibitem{NF-2004}
G. H. Miller, et. \emph{al.}, The National Ignition Facility: enabling fusion ignition for the 21st century, Nuclear Fusion 44 (2004) S228.

\bibitem{POP-2010-Clark}
D. S. Clark, et. \emph{al.}, Plastic ablator capsule design for the National Ignition Facility, Phys. Plasmas 17 (2010) 052703. 

\bibitem{POP-1996}
W. J. Krauser, et. \emph{al.}, Ignition target design and robustness studies for the National Ignition Facility, Phys. Plasmas 3 (1996) 2084.

\bibitem{POP-1998_1953}
D. C. Wilson et. \emph{al.}, The development and advantages of beryllium capsules for the National Ignition Facility, Phys. Plasmas 5 (1998) 1953.

\bibitem{POP-1998_3708}
T. R. Dittrich, et. \emph{al.}, Reduced scale National Ignition Facility capsule design, Phys. Plasmas 5 (1998) 3708.

\bibitem{ramis1988multi}
R. Ramis, R. Schmalz, J. Meyer-Ter-Vehn, {MULTI} - a computer code for
  one-dimensional multigroup radiation hydrodynamics, Comput. Phys. Commun. 49
  (1988) 475.
  
\bibitem{HEDP2019}
G. Mishra, K. Ghosh, Investigation of various methods for wall loss reduction in Inertial Confinement Fusion hohlraums, High Energy Density Physics 33 (2019) 100714.  
 
\bibitem{more1988new}
R. M. More, K. H. Warren, D. A. Young, G. B. Zimmerman, A new quotidian
  equation of state ({QEOS}) for hot dense matter, Phys. Fluids 31 (1988) 3059. 
  
\bibitem{MishraHEDP2018}
G. Mishra, K. Ghosh, A. Ray, N. Gupta, Wide range scaling laws for radiation
  driven shock speed, wall albedo and ablation parameters for high-{Z}
  materials, High Energy Density Physics 27 (2018) 1.
  
\bibitem{Li_POP-2015}
L. Li, et. \emph{al.}, The importance of transmission flux in evaluating the preheat effect in x-ray driven ablation, Phys. Plasmas 22 (2015) 022702.

\bibitem{preheatbarc}
Aeaby C. D., K. Ghosh and G. Mishra, Reduction of preheating in ablators using mid-Z dopants (2020) BARC/ThPS/160/2020.

\bibitem{noneqbarc}
Aeaby C. D., K. Ghosh and G. Mishra, Incorporating non-equilibrium radiation source in MOGS (2020) BARC/ThPS/167/2020. 

\bibitem{Jiang}S. Jiang,  et. \emph{al.}, Influence of Au M-band flux asymmetry on implosion symmetry, Laser and Particle Beams 35 (2017) 337.

\bibitem{Bradley}D. K. Bradley, et. \emph{al.}, Measurements of preheat and shock melting in Be ablators during the first few ns of a NIF ignition drive using the Omega laser. Phys. Plasmas 16 (2009) 042703 .

\bibitem{Hatchett}
S. Hatchett, Ablation gas dynamics of low-Z materials illuminated by soft x-rays,
Tech. rep. Lawrence Livermore National Lab., CA (United States), 1991.
  
\bibitem{Drake}R. P. Drake, High-Energy-Density Physics: Fundamentals, Inertial Fusion and Experimental Astrophysics, Springer, 2006.

\bibitem{Zeldovich}Ya. B. Zeldovich and Yu. P. Raizer, Physics of shock waves and high temperature hydrodynamic phenomena, Academic Press, New York and London, 1967.

\bibitem{Lindl1995}J. Lindl, Development of the indirect-drive approach to inertial confinement fusion 
and the target physics basis for ignition and gain, Phys. Plasmas 2 (1995) 3933.

\bibitem{Saillard}Y. Saillard, Principles of the radiative ablation modeling, Phys. Plasmas 17 (2010) 123302.

\bibitem{Dahmani1}F. Dahmani, Ablation scaling in laser-produced plasmas with laser intensity laser wavelength, and target atomic number, Phys. Fluids B Plasma 4 (1992) 1585.

\bibitem{Dahmani2}F. Dahmani, Experimental scaling laws for mass-ablation rate, ablation pressure in planar laser-produced plasmas with laser intensity, laser wavelength, and target atomic number, J. Appl. Phys. 74 (1993) 622.

\bibitem{Kline}J.L. Kline, J.D. Hager, Aluminum x-ray mass-ablation rate measurements, Matter
Radiat. Extremes 2 (2017) 16.

\end{thebibliography}





\end{document}